\documentclass{aastex62}

\received{xxx}
\revised{yyy}
\accepted{ 05-22-2019}

\begin{document}

\title{A novel approach to study the variability of NGC 5548}

\correspondingauthor{A. Bewketu Belete}
\email{asnakew@fisica.ufrn.br}

\author{A. Bewketu Belete}
\altaffiliation{Departamento de F\'isica Te\'orica e Experimental, Universidade Federal do Rio Grande do Norte, Natal, RN 59078-970, Brazil}
\affiliation{Departamento de F\'isica Te\'orica e Experimental, Universidade Federal do Rio Grande do Norte, Natal, RN 59078-970, Brazil}
\author{L. J. Goicoechea}
\affiliation{Departamento de Física Moderna, Universidad de Cantabria (UC), Avda. de Los Castros s/n, 39005 Santander, Spain}
\author{I. C. Le\~ao}
\affiliation{Departamento de F\'isica Te\'orica e Experimental, Universidade Federal do Rio Grande do Norte, Natal, RN 59078-970, Brazil}
\author{B. L. Canto Martins}
\affiliation{Departamento de F\'isica Te\'orica e Experimental, Universidade Federal do Rio Grande do Norte, Natal, RN 59078-970, Brazil}
\author{J. R. De Medeiros}
\affiliation{Departamento de F\'isica Te\'orica e Experimental, Universidade Federal do Rio Grande do Norte, Natal, RN 59078-970, Brazil}



\begin{abstract}
Understanding the properties of the continuum radiation and broad emission lines of active galactic nuclei provide significant information not only to model the radiation mechanism and constrain the geometry and kinematics of the broad-line region (BLR) but also to probe the central engine of the sources. Here we investigate the multifractal behaviour of the H$\beta$ emission line and the 5100 {\AA} continuum flux light curves of NGC 5548. The aim is to search for multi-scaling signatures in the light curves and check if there is a possible nonlinear relationship between them. To this end, we use a multifractality analysis technique called Multifractal Detrended Moving Average (MFDMA) analysis. We detect multifractal (nonlinear) signatures in the full monitoring and densely sampled period of the H$\beta$ line and 5100 {\AA} continuum light curves of NGC 5548, possibly indicating the presence of complex and nonlinear interaction in the 5100 {\AA} continuum and H$\beta$ emission line regions. Moreover, the degree of multifractality of H$\beta$ line is found to be about twice that of the 5100 {\AA} continuum. The nonlinearity of both emissions could be generated when the broad-line region reprocesses the radiation from the central compact source. Finally, we found that anti-persistent long-range temporal correlation is the main source of the multifractality detected in both light curves.
\end{abstract}



\section{Introduction} \label{sec:intro}

In the study of active galactic nuclei (hereinafter AGNs) understanding the nature of the central engine, the geometry and kinematics of the broad-line region (hereinafter BLR), and the connection between the continuum and the BLR remain the most important questions. Several studies to uncover the working mechanism of the central engine (eg., \citealt{doi:10.1111/j.1365-2966.2008.13951.x, 0004-637X-686-2-892, 2007ASPC..373..755N}), to study the variability and constrain the geometry and kinematics of the BLR (e.g., \citealt{2017A&A...607A..32B, 2017PASA...34...42Y, 10.3389/fspas.2017.00018, 2012ApJ...747...62C, doi:10.1111/j.1365-2966.2011.19463.x, 2017ApJ...849..146G, 2013ApJ...764...47G, 2009NewAR..53..140G, 2009ApJ...704L..80D, doi:10.1146/annurev.astro.38.1.521, 1996ApJ...466..704C}), and to learn the correlation between the continuum and the BLR (e.g., \citealt{2018ApJ...869..143P, 2005ApJ...629...61K, 1980ApJ...241..894Y}) have been published. It is now believed that AGNs are powered by mass accretion onto a super-massive black hole (e.g., \citealt{2011A&A...532A.102R, 2004MNRAS.351..169M, 1998ApJ...498..570D} and the references therein). Most, if not all, AGNs consist of mainly four important regions such as the central engine (super-massive black hole), the continuum region, the BLR, and the narrow-line region (NLR). The study of broad-emission lines of AGNs is very important not only to constrain the structure and kinematics of the BLR but also to understand the nature of the central continuum and the engine powering these sources. The broad emission lines we observe in the UV/optical regime are generated by photoionization in the outer regions of an accretion disk (hereinafter AD) that surrounds the central black hole (\citealt{kollatschny2013accretion}). Since the BLR is close to the center of the AGN, it responds to continuum variability. It has been recognized that knowledge about the variations in the broad emission lines of quasars provides us valuable information that can be used to constrain the size of the BLR and the continuum region (e.g., \citealt{doi:10.1111/j.1365-2966.2004.07349.x}). Furthermore, the correlations between the continuum and emission-line properties of AGNs gives the sizes of line-emitting regions and provide a potentially valuable information to probe the unobservable ionizing continuum region and central engine in AGNs (\citealt{1986ApJ...305..175G, 1973ApL....13..165C}). The technique based on the lag between the intrinsic variability of the continuum and of the broad emission lines, reverberation mapping, has been applied to date in studying the geometry and kinematics of the BLR  (e.g., \citealt{10.3389/fspas.2017.00071, 2006plco.book...89P, 1993PASP..105..247P, Article, article}). Though it  has been applied largely following its success, reverberation mapping technique of determining the size of BLR remains with several limitations (e.g., \citealt{2013ApJ...764..160G} and the references therein). \\
Multifractality analysis has been applied to different AGN light curves (e.g., see \citealt{2019arXiv190205470B, 10.1093/mnras/stz203, doi:10.1093/mnras/sty1316}), and here we study the  multifractal, and thus the nonlinear, behaviour of the H$\beta$ emission line and 5100 {\AA} continuum flux of  the Seyfert type I galaxy NGC 5548. Type I Seyfert galaxies show broad emission lines because their BLR is not hidden by a dusty torus (e.g., see \citealt{2015A&A...576A..73P, 1998ApJ...498..570D}, and references therein). Significant variations in the intensities of the broad emission lines are identified as one of the features of Seyfert I galaxies (\citealt{1990MNRAS.244..138R} and the references therein), and \citet{1973ATsir.777....1L} also showed that the optical continuum of NGC 5548 is highly variable. NGC 5548 was among the first Seyfert galaxies to be studied (\citealt{2016ApJS..225...29B}) and is one of the nearby active galaxies that have deserved more attention (\citealt{2015A&A...575A..22M, kollatschny2013accretion}). In recent years, spectral variability and other properties of this object has been intensively studied by the Space Telescope and Optical Reverberation Mapping campaign (e.g., \citealt{2017ApJ...846...55M, 2017ApJ...835...65S, 2017ApJ...837..131P, 2016ApJ...824...11G, 2016ApJ...821...56F, 2015ApJ...806..128D, 2015ApJ...806..129E}).\\ 
The first reverberation mapping of NGC 5548 (\citealt{1986AJ.....92..552P}) has shown that H$\beta$ responded to continuum changes with a delay of only a few weeks. According to \citet{2007ApJ...668..708S}, inspection of individual broad H$\beta$ profiles over a 30-year period reveals that the broad emission line profiles can undergo dramatic changes (from a typical single-peaked profile centered near the systemic redshift of the galaxy, to profiles that show prominent blue/red peaks or two peaks). \citet {2003ApJ...587..123G} have found a nonlinear relationship between the H$\beta$ emission line and the optical continuum flux of NGC 5548. \citet{2009ApJ...704L..80D} have seen evidence for out-flowing, in-falling, and virialized BLR gas motions based on the optical continuum and broad H$\beta$ emission line variations observed in the nuclear regions of selected Seyfert galaxies including NGC 5548. Using the 43 years (1972 to 2015) monitoring data of NGC 5548, (\citealt{2016ApJS..225...29B}) revealed $\sim$ 5700 day periodicity in the continuum light curve, the H$\beta$ light curve, and the radial velocity curve of the red-wing of the H$\beta$ line. Orbiting dusty and dust-free clouds, a binary black hole system, tidal disruption events, and the effect of an orbiting star periodically passing through an AD are explained to be the possible physical mechanisms for the periodicity. Additional studies of NGC 5548 were performed by \citet{article, 2015A&A...575A..22M, 2009ApJ...705..199B, 2007ApJ...662..205B, 1999ApJ...510..659P, 1992ApJ...392..470P, 1991ApJ...368..119P, 1990ApJ...353..108N}.\\
Our aim is to characterize the multifractal (nonlinear) behaviour of the H$\beta$ emission line and the 5100 {\AA} continuum flux of NGC 5548 and search for any possible nonlinear relationship, or similarity/difference in the degree of nonlinearity between them - how emission line response to continuum variations?. This provides us a potentially valuable information to model the structure and dynamics of both the continuum and its associated broad-line regions, which in turn can be used as an input to better understand the physics of the central engine and accretion.
This work is organized as follows. Section \ref{data} includes the data, and analysis method we used. The results are given in section \ref{res}. Section \ref{dis} presents the discussion of the results. Finally, the main conclusions are summarized in section \ref{concl}.

\section{OBSERVATIONAL DATA AND ANALYSIS METHOD}\label{data}
\subsection{ Light curves}
\begin{figure*}
\centering
\includegraphics[scale=0.7]{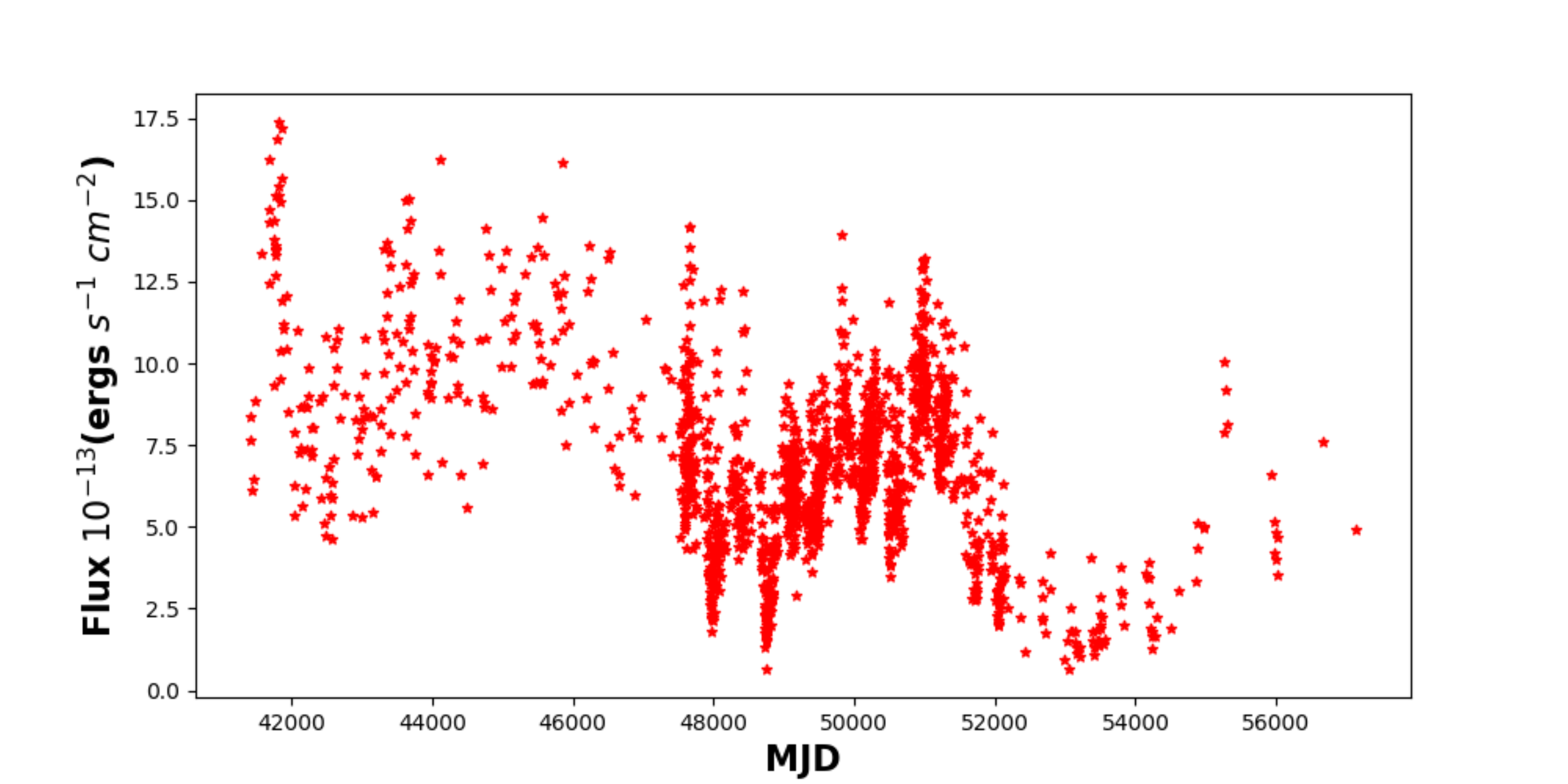}
\includegraphics[scale=0.7]{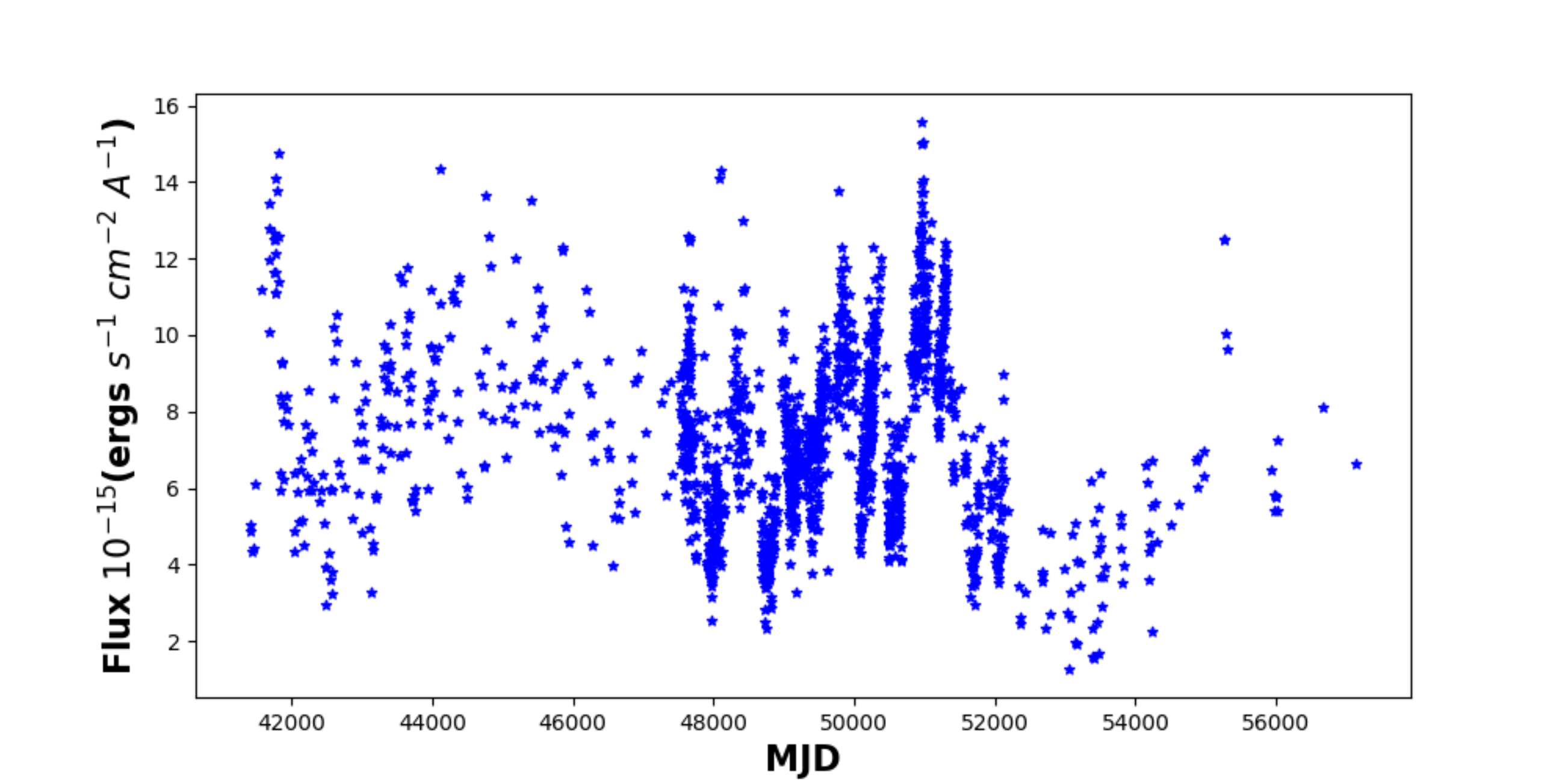}
\caption{The light curves of the H$\beta$ emission line (top) and the 5100 {\AA} continuum flux (bottom) of NGC 5548.}
\label{fig1}
\end{figure*}

We use the log-term (1972 to 2015) light curves of the H$\beta$ emission line and 5100 {\AA} continuum flux of the Seyfert type I galaxy NGC 5548\footnote{http://vizier.cfa.harvard.edu/viz-bin/VizieR-3?-source=J/ApJS/225/29/table2} (see Fig. \ref{fig1}; \citealt{2016ApJS..225...29B}).

\subsection{Framework for analysing light curves}\label{sec2}
In the last few decades, stochastic processes have been used to describe the variability of quasars, e.g., computing structure functions ($SFs$) or simulating light curves. A popular model to account for quasar variability, mainly in the optical band, is the so-called damped random walk (DRW; e.g., \citealt{2017ApJ...847..132G, 10.1093/mnras/stv1230, 2014IAUS..304..395I, 2013ApJ...765..106Z, 2012ApJ...753..106M, 2011ApJ...735...80Z, 2010ApJ...708..927K}). The $SF$ for the DRW model is given by (\citealt {2010ApJ...721.1014M}, and reference therein):
\begin{equation}\label{eq1}
SF(\Delta t) = SF{\infty}(1-e^{-|\Delta t|/\tau})^{1/2}
\end{equation}
where $\tau$ is the characteristic time, $SF{\infty}$ = $\sqrt 2$$\sigma$, and $\sigma$ is the variability amplitude.  This $SF$ has two asymptotic
regimes: 
\begin{equation}\label{eq2}
SF(\Delta t >>\tau) \equiv  SF{\infty},~
SF(\Delta t <<\tau) = SF{\infty}\sqrt {\frac{|\Delta t|}{\tau}}.
\end{equation}
The characteristic timescale $\tau$ and $SF${$\infty$} are the two key parameters, and a behaviour of $SF$ $\propto$ 
$|\Delta t|^{\beta}$  is related to a power spectral distribution $PSD$ $\propto$ $f^{-\delta}$, where $\delta$ = 2$\beta$+1. Thus, in Eq.  \ref{eq2}, one deals with the standard asymptotic limit of a random walk at short values of $\Delta t$, i.e., a Wiener process with $\delta$ = 2, while the damped nature manifests at long values of $\Delta t$, when the $SF$ and $PSD$ become constant ($\delta$ = 0). In addition, the power-law index $\delta$ can be also related to the generalized Hurst exponent from a multifractal analysis (see discussion below Eq. \ref{eq9}). This allows us to compare multifractality-based power-law indexes and those predicted by the DRW model.\\
Naturally, complex systems generate time series that exhibit fluctuations over a range of time scales and/or broad distributions of the values (\citealt{kantelhardt2009fractal}). These natural fluctuations are often found to follow a scaling relation over several orders of magnitude, from which one can characterize the time series and the generating complex systems by a set of scaling exponents (fractals). This characterization of the time series and the systems can be used to make comparison with other systems or/and models. Moreover, determining the nature of scaling between fluctuations in a variable of interest and whether some type of power-law exponents present for various statistical moments at different scales or not is very significant to better characterize the behaviour of the time series and the system generating the time series. Since the dynamics of such complex systems cannot be described by a single power exponent (monofractal), it is necessary to apply a different formalism that can explore the set of power exponents (fractals/multifractals) to fully characterize the systems. Particularly, a multifractal analysis approach is applicable when different scaling exponents are required to describe different parts of the series. There are different techniques that have been developed to characterize the properties of stationary and non-stationary fractal and multifractal time series. Some of them are Hurst's rescaled-range analysis (R/S), Fluctuation analysis (FA), Autocorrelation Function (ACF) analysis, Spectral Analysis, wavelet transform module maxima (WTMM) method, detrended fluctuation analysis (DFA), multifractal detrended fluctuation analysis (MFDFA), detrending moving average (DMA), and Multifractal Detrended Moving Average (MFDMA) algorithm (\citealt{2010PhRvE..82a1136G, kantelhardt2009fractal}, and the references therein).\\
The standard fluctuation analysis (FA) is a fractal analysis technique for stationary time series and is based on random walk theory (\citealt{kantelhardt2009fractal}, and the references therein). For example, for a time series $x(t), t = 1, 2, 3, ..., N$, the global profile is given by eq. \ref{eq3} and study how the fluctuations of the profile, in a given segment size $s$, increase with $s$. One can determine the square-fluctuations of the profile scale with $s$, by dividing each record of $N$ elements into $N_{s} = int(N/s)$ non-overlapping segments of size $s$ starting from the beginning and another $N_{s}$ non-overlapping segments of size $s$ starting from the end of the considered series. Further more, the mean fluctuation function can also be estimated by averaging the square-fluctuations over all sub-sequences (see \citealt{kantelhardt2009fractal} for the detail). Since FA is unable to analyze non-stationary fractal time series, a different technique called DFA was introduced by \citet{PhysRevE.49.1685}. The establishment of DFA was basically to detect long-range (auto-) correlations in non-stationary time series. The formalism DFA is also based on random walk theory and basically represents a linear detrending version of FA (\citealt{kantelhardt2009fractal}. But, the technique DFA presents abrupt jumps in the detrended profile, which is the limitation of the method. To fix this drawback several modifications and extensions have been introduced. DMA is one of these modifications. Further more, DMA was extended to MFDMA to analyze multifractal time series and multifractal surfaces (\citealt{2010PhRvE..82a1136G}).\\
Here we apply the backward ($\theta = 0$) one-dimensional MFDMA algorithm briefly discussed in \citet{2010PhRvE..82a1136G}. 
In multifractality analysis the most crucial parameters to describe the structural behaviour of a time series \textit{x(t)} are (i) a $q^{th}$-order fluctuation function $F_q$($n$), from which one can calculate the local Hurst exponent,$ h$($q$) (ii) the multifractal scaling exponent function $\tau$($q$), and (iii) the multifractal spectrum function \textit{f}(\textit{$\alpha$}). We obtain these parameters following the procedures explained in \citet{2010PhRvE..82a1136G}: \\
1. Consider a time series $x(t),t = 1, 2, 3, ..., N$. The time series is then reconstructed as a sequence of cumulative sums given by:
\begin{equation}\label{eq3}
y(t)=\sum_{i=1}^t x(i),   \quad  t=1,2,3,...,N,                 
\end{equation}
where $N$ is the length of the data.\\
 2. We calculate the moving average function $\tilde{y}(t)$ of Eq.~\ref{eq1} in a moving window using the relation 
\begin{equation}\label{eq4}
\tilde{y}(t)={1\over n}\sum_{k=-\lfloor(n-1)\theta \rfloor}^{\lceil(n-1)(n-\theta)\rceil}y(t-k),
\end{equation}
where $n$ is the segment (window) size. The lowest segment size ($n_{min}$) is chosen to be 10 and the maximum ($n_{max}$) is around 10 \% of $T$, where $T$ is the length of the time series. In addition, for our case, $\theta$ is adopted as zero, referring to the backward moving average. See \citealt{ doi:10.1093/mnras/sty1316, 2010PhRvE..82a1136G} for detailed explanation.\\
3. We remove the trend from the reconstructed time series $y(t)$ using the function $\tilde{y}(t)$ and the residual sequence $\epsilon(t) as follows$:
\begin{equation}\label{eq5}
\epsilon(i)= y(i)-\tilde{y}(i),
\end{equation}
where $n-\lfloor(n-1)\theta\rfloor\leq{i}\leq{N}-\lfloor(n-1)\theta\rfloor$. The residual time series $\epsilon(t)$ is subdivided into ${N_n}$ disjoint segments with the same size $n$ given by $N_{n} = \lfloor{N\over n}-1\rfloor$. In this sense, the residual sequence $\epsilon(t)$ for each segment is denoted by $\epsilon_v$, where $\epsilon_v(i)=\epsilon(l+1)$ for $1\leq{i}\leq{n}$ and $l=(v-1)n$.\\
4. Then we calculate the root-mean-square function using the relation:
\begin{equation}\label{eq6}
{F_v}(n)= \left\{{1\over n}\sum_{i=1}^n {\epsilon_v}^2(i)\right\}^{1/2}.
\end{equation}
5. We determine the overall fluctuation function $F_q (n)$ as:
\begin{equation}\label{eq7}
{F_q}(n)= \left\{{1\over {N_n}}\sum_{v=1}^{N_n} {F_v}^2(n)\right\}^{1/q}
\end{equation}
for all $q\neq0$, where $q$ is the statistical moment or power argument which takes a set of zero-mean numbers, for our case it is chosen to be in the interval [-5 5].\\ When $q=0$, according to L'H\^{o}spital's rule, we have:
\begin{equation}\label{eq8}
{F_{0}}(n)= exp\left\{{1\over 2{N_n}}\sum_{v=1}^{N_n}ln[{F_v}^2(n)]\right\},
\end{equation}
 6. We determine the power-law relation between the overall fluctuation function $F_q (n)$ and the segment size $n$ by using the relation:
\begin{equation}\label{eq9}
F_q(n)=n^{h(q)},
\end{equation}
 where $h$($q$) is the slope (the scaling exponent), to be calculated from the log ($F_q$($n$)) versus log($n$) plot using the least square fitting technique. This relationship is applicable if the time series has long-range power-law correlations, or if $F_q$($n$) increases for larger values of $n$ as a power-law. For a monofractal series, a series with single exponent, the Hurst exponent (\citealt{1951Hurst}) is sufficient to characterize the behaviour of the series at different scales, and thus $H$ = $h$($q$), where $H$ is the generalized Hurst exponent. For a multifractal time series, the time series scales differently at different time scales producing a range of exponents, and thus $H$ is not sufficient to describe the behaviour of the series. Therefore, for a multifractal time series H $\neq$ $h$($q$), and the scaling exponent $h$($q$) can be thought of as ''local Hurst exponents'' instead (\citealt{Hampson:11}). The local Hurst exponents defines a continuum between a noise like time series and a random walk like time series. For a stationary time series, such as fractional Gaussian noise (fGn), $h$($q$ = 2) will be between 0 and 1, and $h(2)$ = $H$ (\citealt{2006JSMTE..02..003S, 1988Feder}). For a nonstationary time series (i.e., a time series with time-dependent variance) with a random walk or fractional Brownian motion (fBm), the local Hurst exponent $h$($q$) $>$ 1 (\citealt{ihlen2012introduction, 2006JSMTE..02..003S}), and related to $H$ as $H$ = $h$($q$ = 2) $-$ 1 (\citealt{Hu_2009, 2006JSMTE..02..003S}).  The $H$ values 0$<$ $H$ $<$ 0.5 and 0.5 $<$ $H$ $<$ 1 indicate anti-persistence and persistence long-range correlations, respectively, whereas, $H$ = 0.5 indicates uncorrelated time series. The temporal or spatial fluctuation of complex systems can be characterized by a power-law decaying power spectral density $PSD$ $\propto$ $f^{-\delta}$, where $\delta$ is power spectrum scaling exponent. According to random fractal theory, $\delta$ can be related to the generalized Hurst exponent $H$ as $\delta = 2H+1$ (\citealt{Hu_2009, 2006JSMTE..02..003S}). For a fGn process, $\delta$ $<$ 1, whereas, $\delta$ $>$ 1 for fBm (\citealt{Hampson:11}, and reference therein).\\

 7. Using the calculated $h$($q$), we determine the Renyi scaling exponent $\tau$($q$)  as follows: 
\begin{equation}\label{eq10}
\tau(q)=qh(q)-D_f,
\end{equation}
where $D_f$ is the fractal dimension of the geometric support of the multifractal measure. For our case, $D_f = 1$, since we are applying the MFDMA for one-dimensional time series analysis.\\
8. We determine the singularity strength function (the H{\"o}lder exponent) $\alpha$($q$), and finally estimate the multifractality spectrum function $f$($\alpha$) as follows:
\begin{equation}\label{eq11}
\alpha(q)=d\tau(q)/dq,       
\end{equation}
and
\begin{equation}\label{eq12}
f(\alpha)=q\alpha-\tau(q). 
\end{equation}


The $\Delta\alpha = \alpha_{max} - \alpha_{min}$ (the range of singularity strength $\alpha$) is the measure of the range of multifractal singularity strength and denotes the range of exponents present in a time series. If there is strong nonlinearity in the scaling exponent curve, i.e., different slope for the negative and positive $q$ values, then we can have wider $\Delta\alpha$, which reflects strong multifractality behaviour in the data considered (\citealt{2004PCE....29..295T, kantelhardt2002multifractal}). A physical system is said to be intermittent if it has a wide multifractal spectrum or concentrates into a small-scale features with large magnitude of fluctuations enclosed by extended areas of less strong fluctuations \citep{monin2007statistical, 1995tlan.book.....F, moffatt_1994}. In addition to the width, the  symmetry in the shape of $\alpha$, defined as $A = (\alpha_{max}-\alpha_{0})/(\alpha_{0}-\alpha_{min})$, where  $\alpha_{0}$ is the value of  $\alpha$ when $f(\alpha)$ assumes its maximum value, provides valuable information about the singularities present in the time series. The asymmetry presents three shapes: asymmetry to the right-skewed ($A >1$), left-skewed ($0< A <1$), or symmetric ($A = 1$). A brief explanation of this can be found in \citet{doi:10.1093/mnras/sty1316} and the references therein. 
\section{Results}\label{res}
\begin{figure*}
\centering
\includegraphics[scale=0.6]{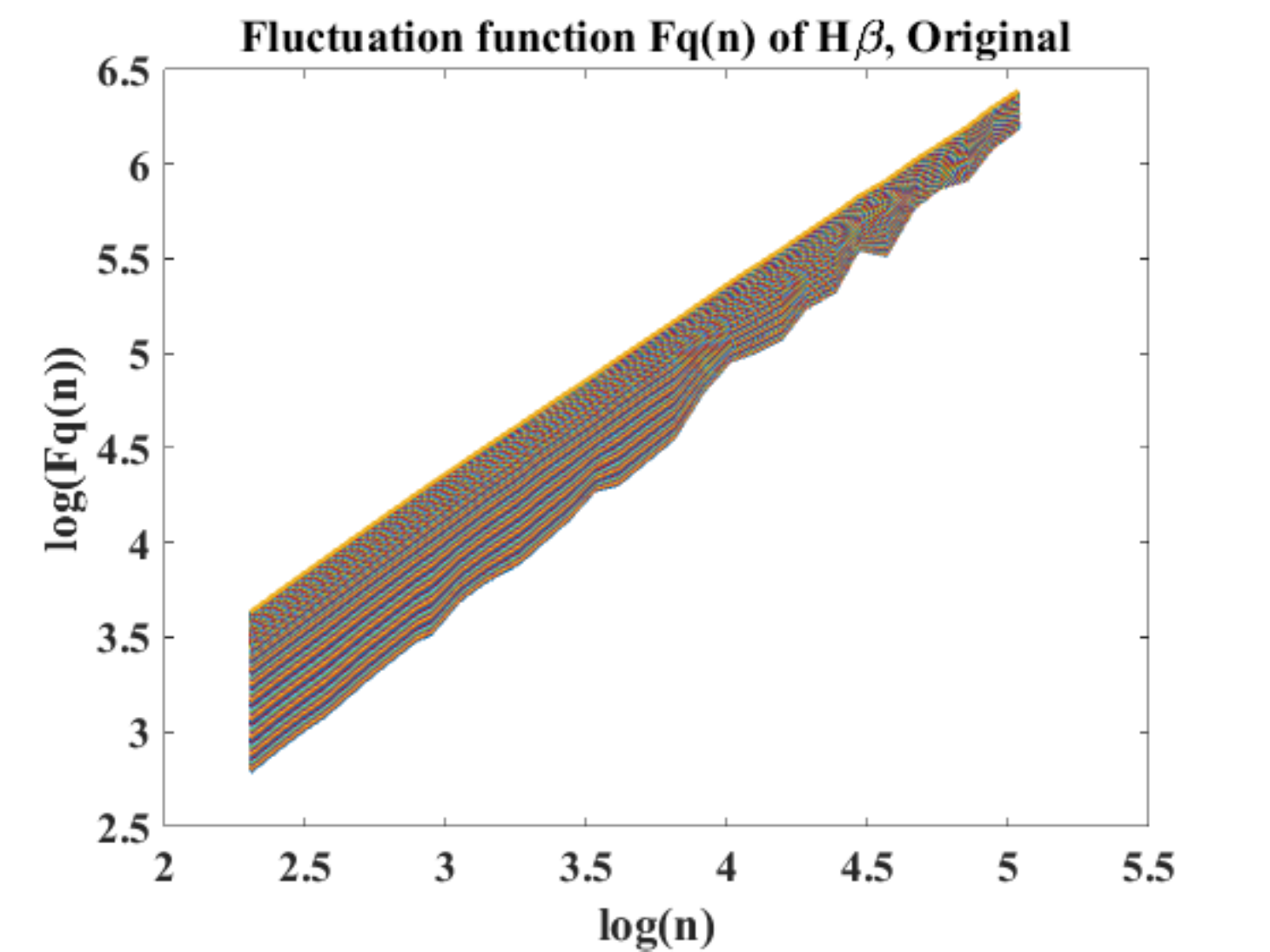}
\includegraphics[scale=0.6]{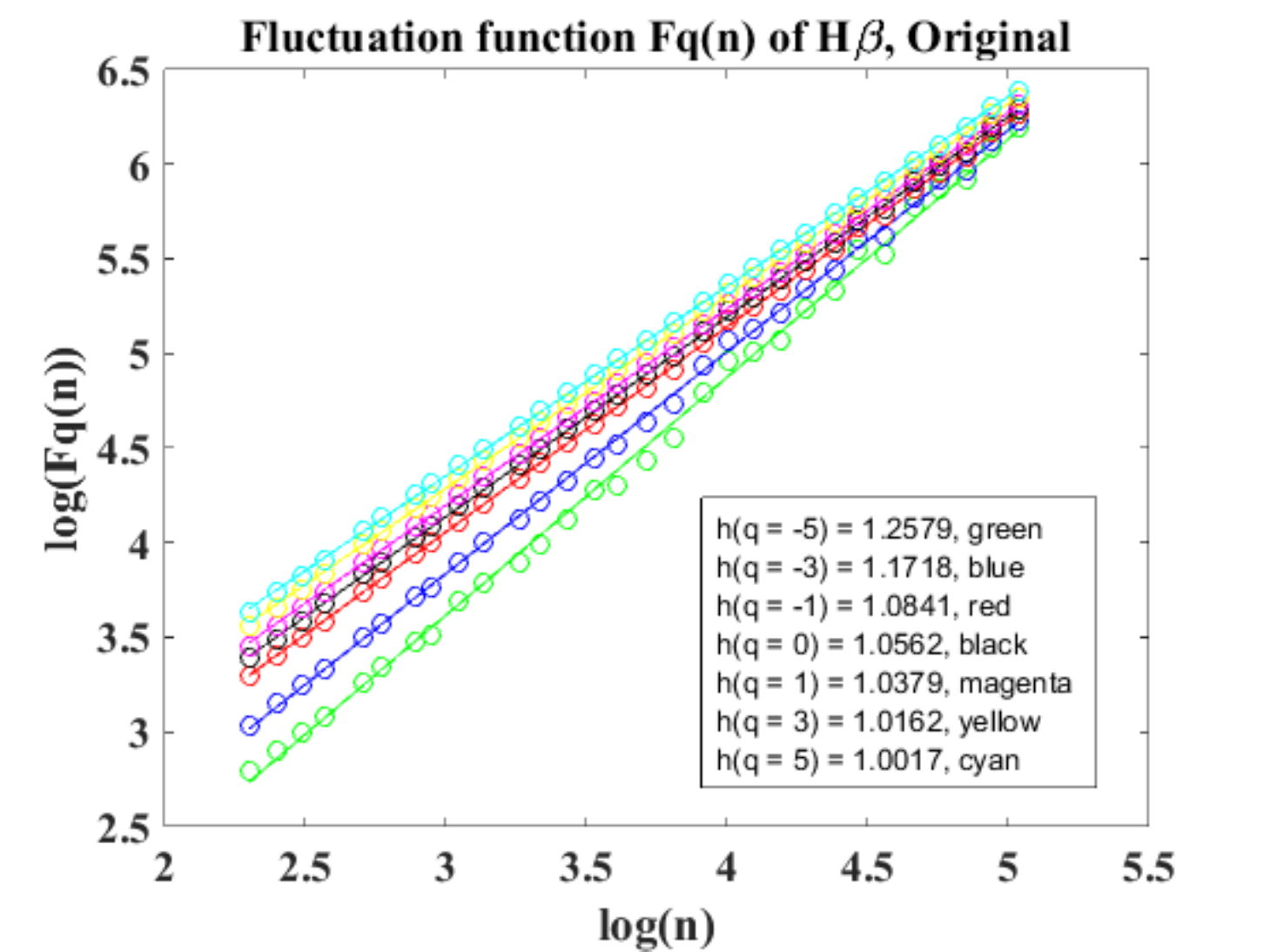}
\includegraphics[scale=0.6]{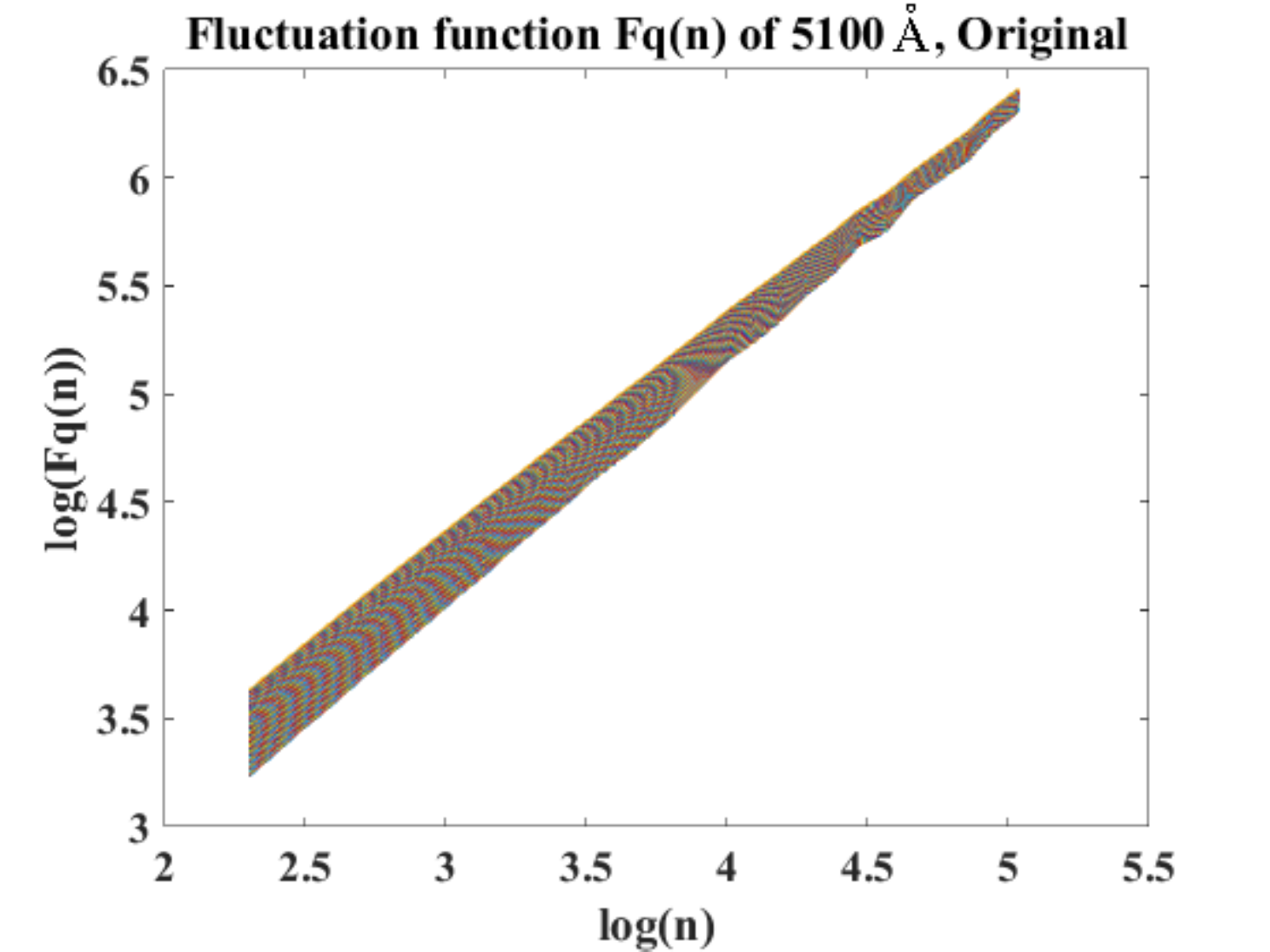}
\includegraphics[scale=0.6]{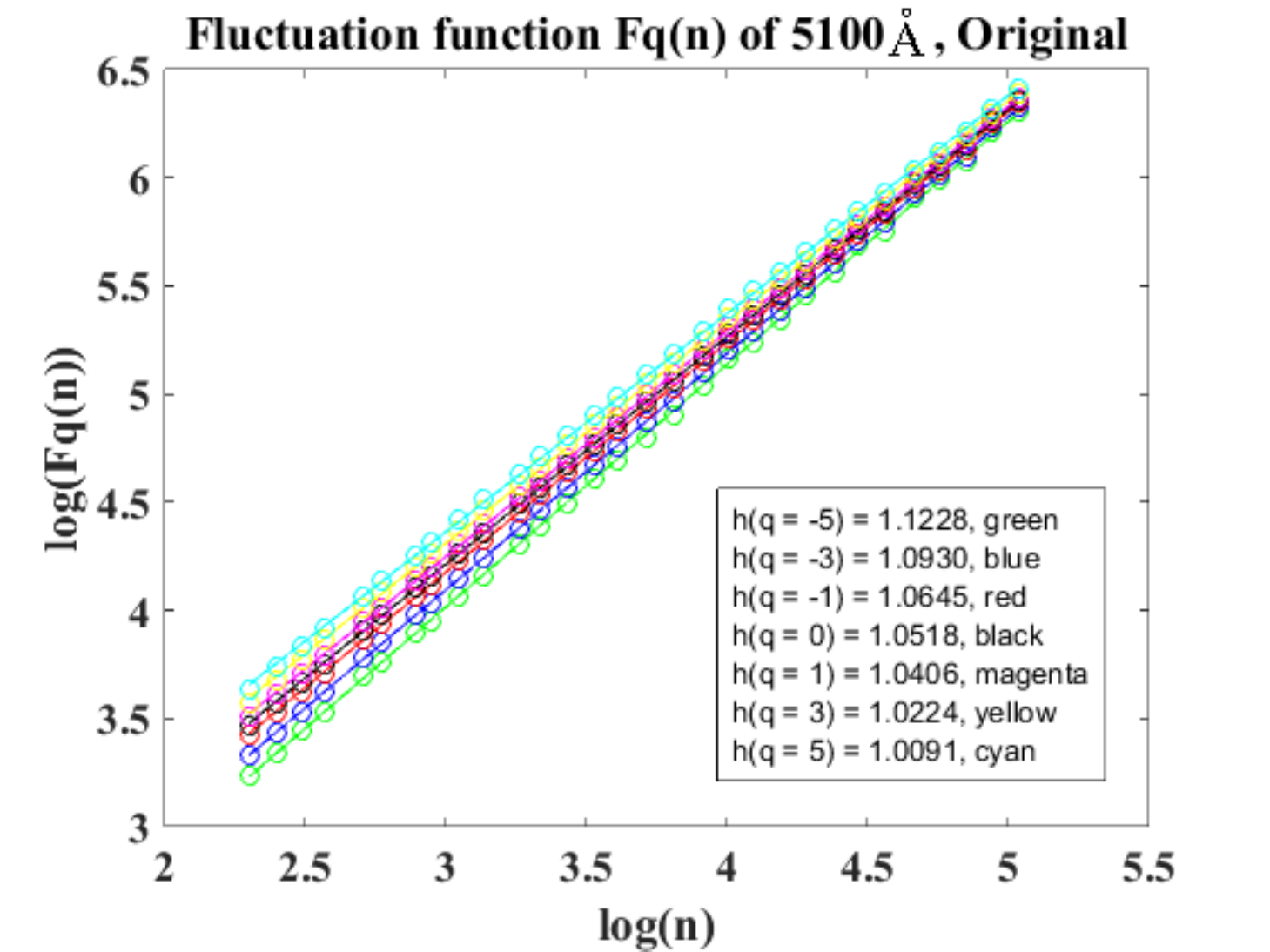}
\caption{Top panel: the fluctuation function $F_q$($n$) of the original light curve of H$\beta$ emission line. The one on right side is after least square fitting for selected $q$ values. Bottom panel: the same as the top panel but for the 5100 {\AA} continuum flux.}
\label{fig2}
\end{figure*}

\begin{figure*}
\centering
\includegraphics[scale=0.6]{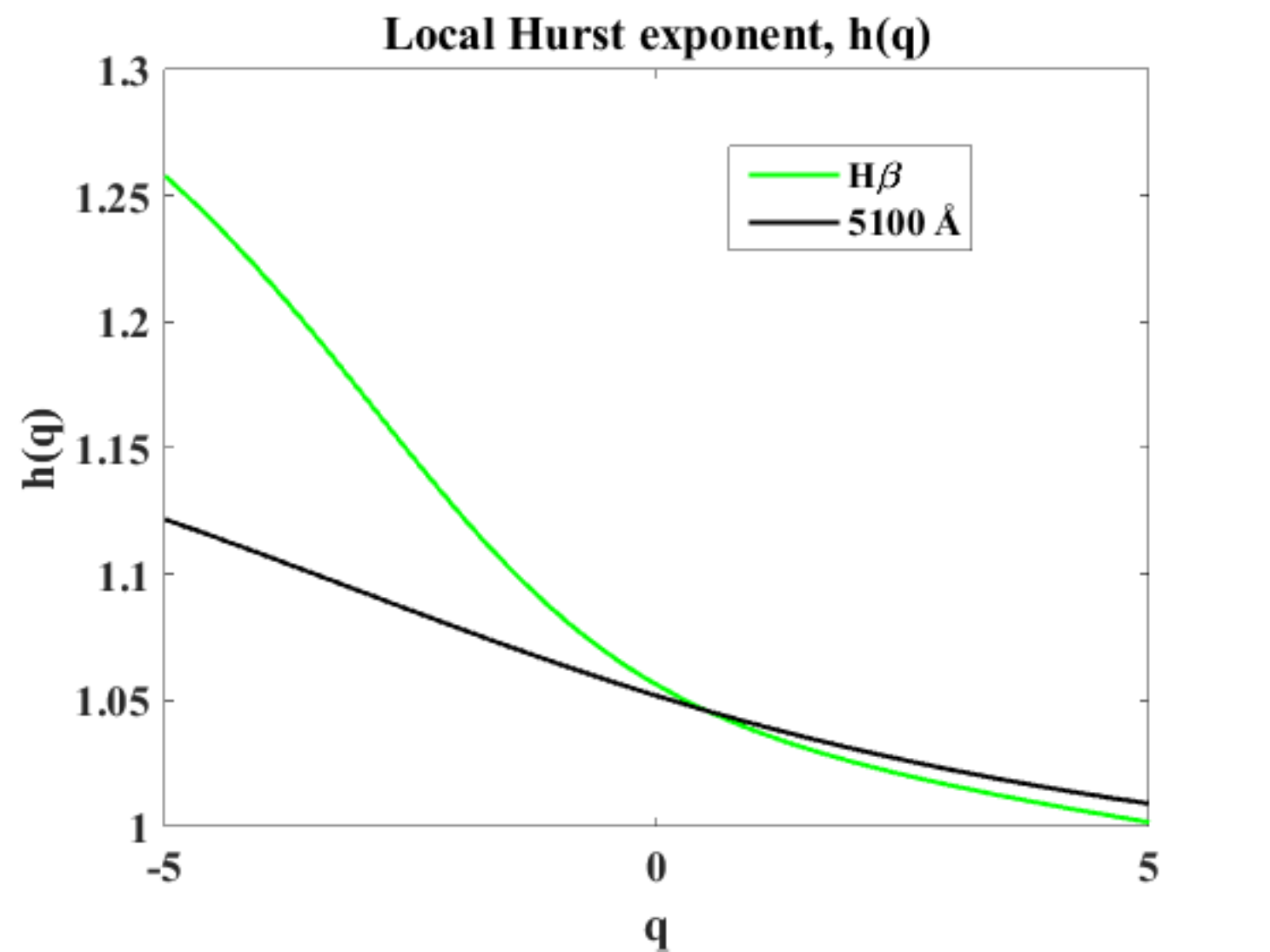}
\includegraphics[scale=0.6]{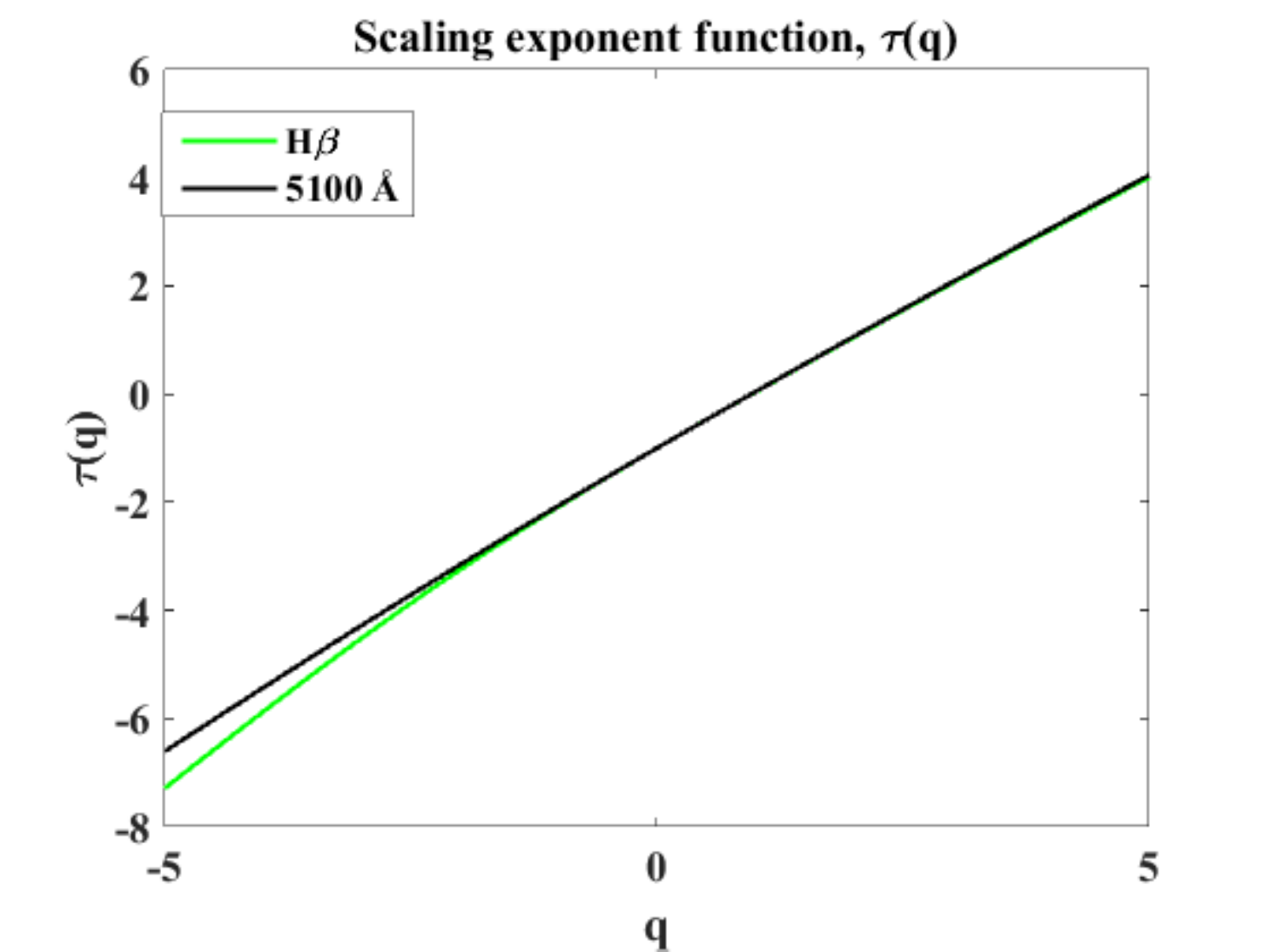}
\includegraphics[scale=0.6]{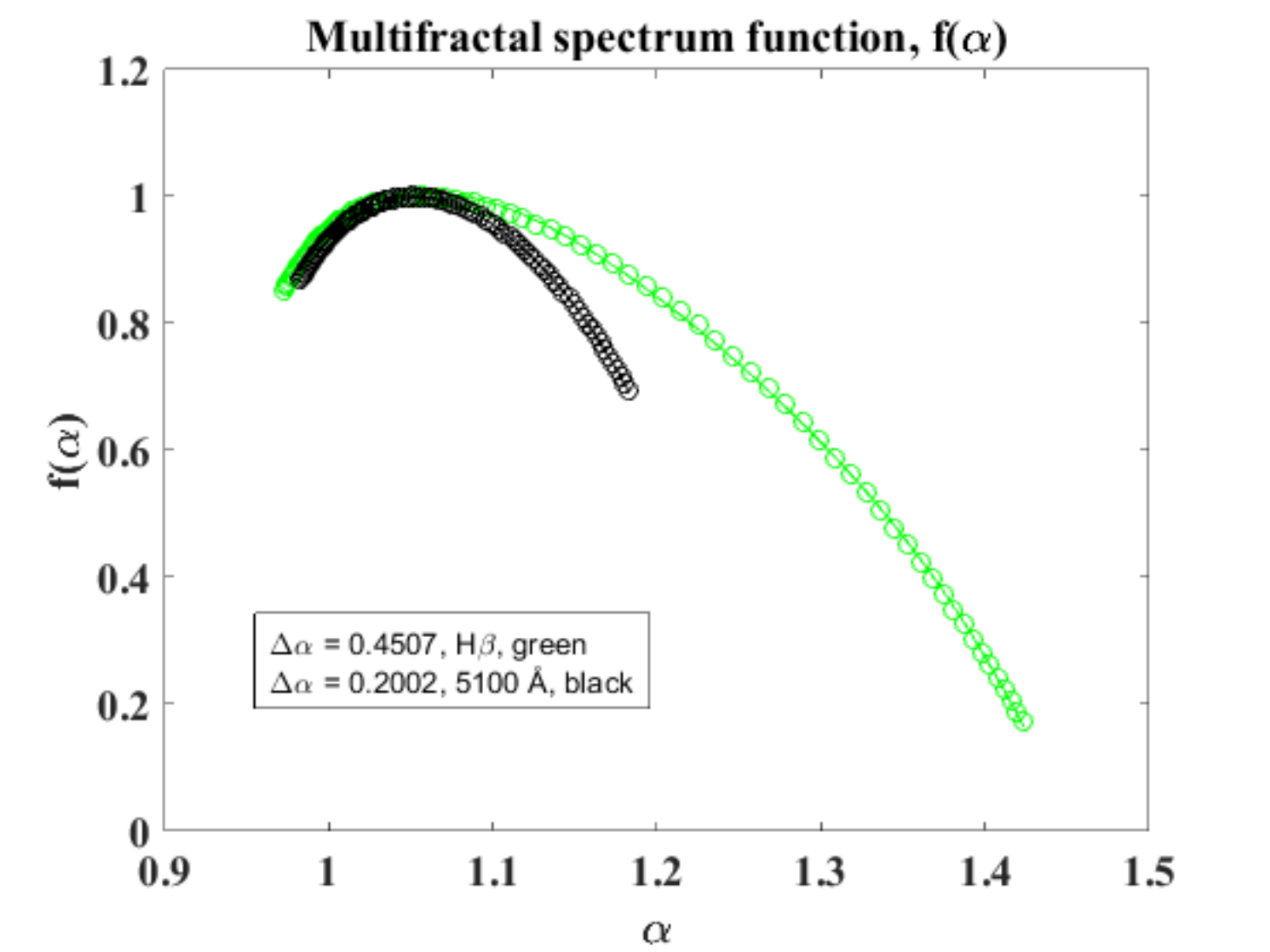}
\caption{Top panel: the local Hurst exponent $h$($q$) (left) and the scaling exponent function $\tau$($q$) (right) for the H$\beta$ emission line (green) and the 5100 {\AA} continuum flux (black). Bottom panel: the multifractal spectrum function $f$($\alpha$) for the H$\beta$ emission line (green) and the 5100 {\AA} continuum flux (black). The circles represent the data points, whereas, the lines represent the fit to fourth-order polynomial.}
\label{fig3}
\end{figure*}

\begin{figure*}
\centering
\includegraphics[scale=0.6]{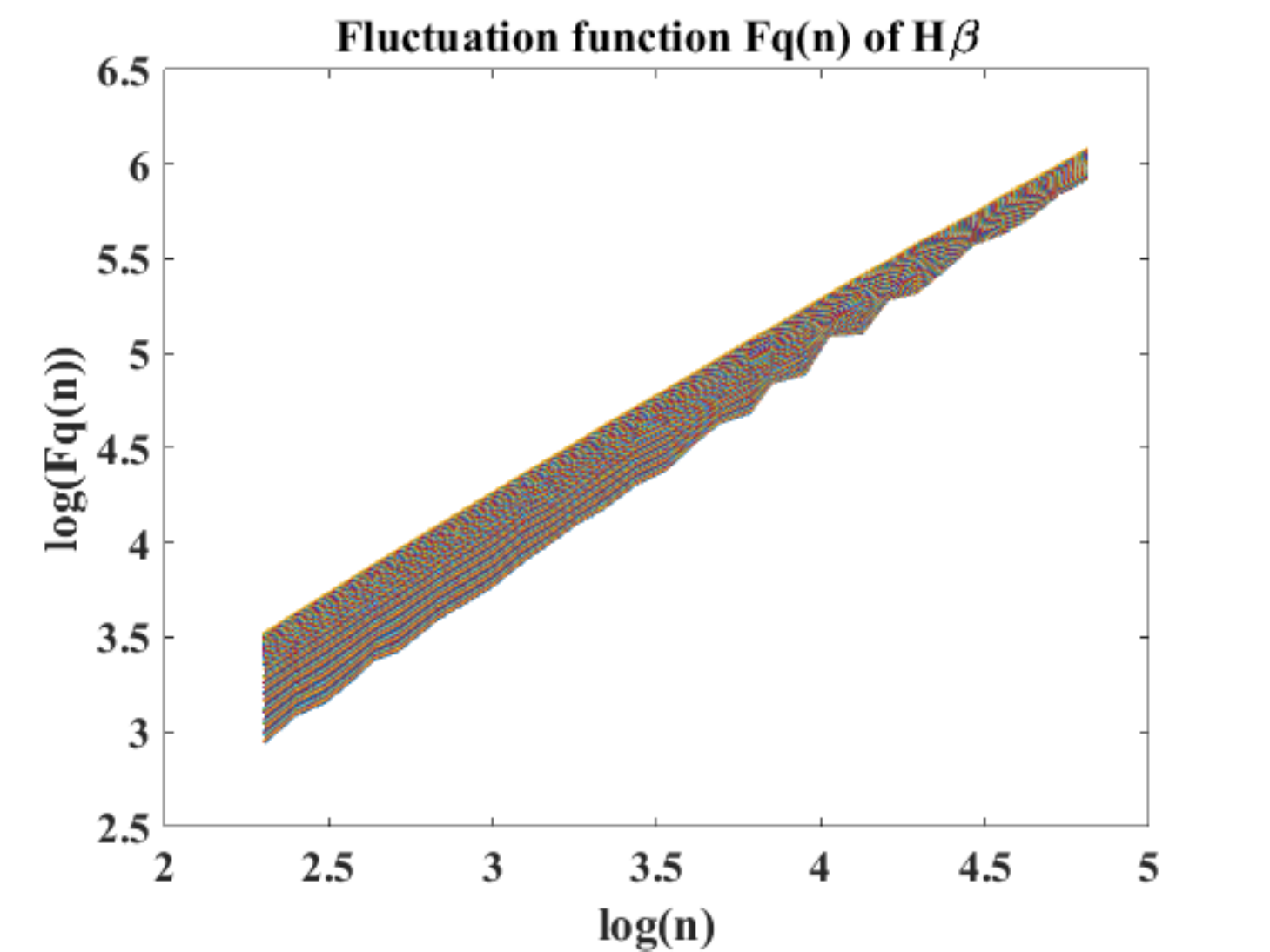}
\includegraphics[scale=0.6]{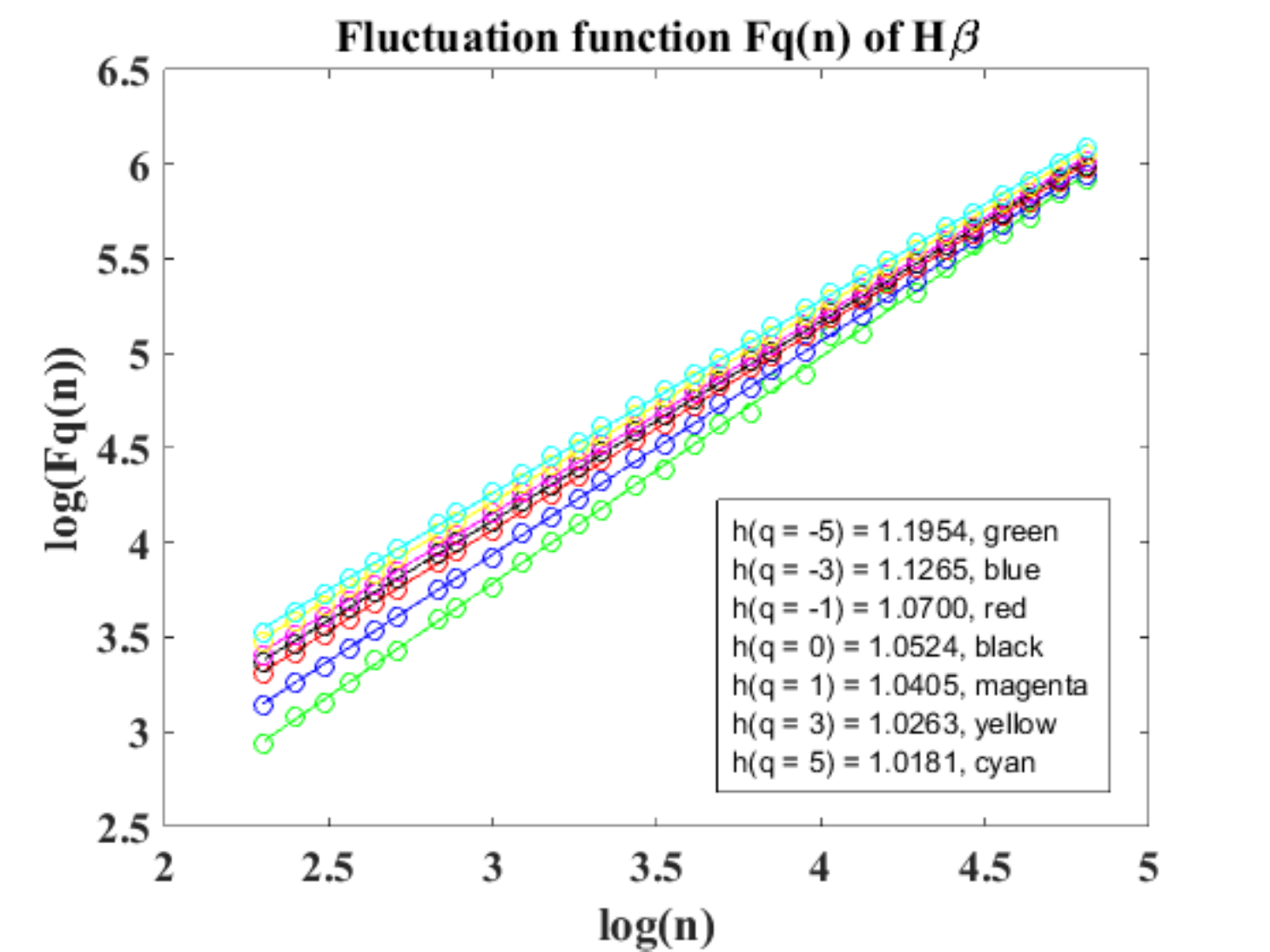}
\includegraphics[scale=0.6]{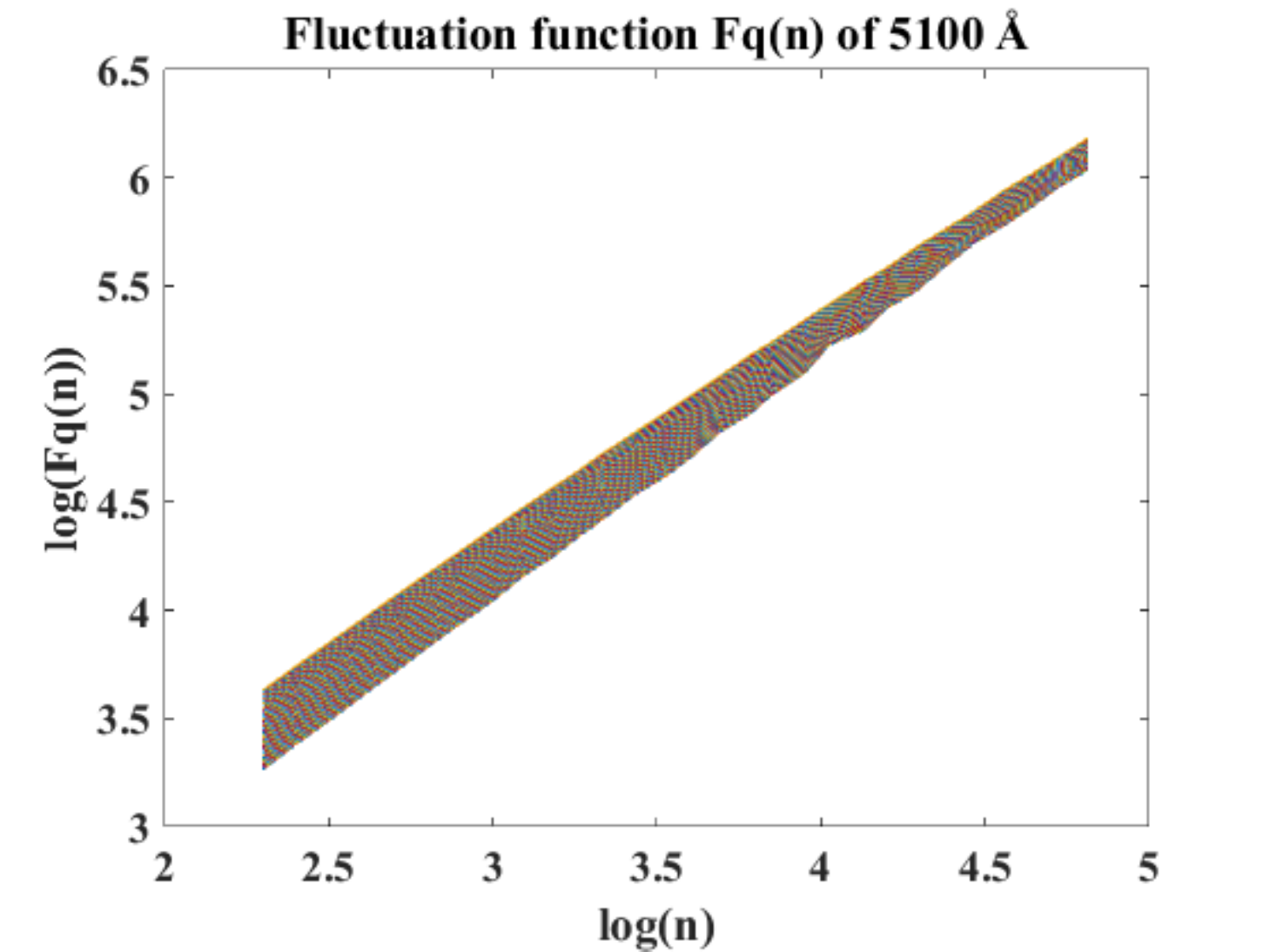}
\includegraphics[scale=0.6]{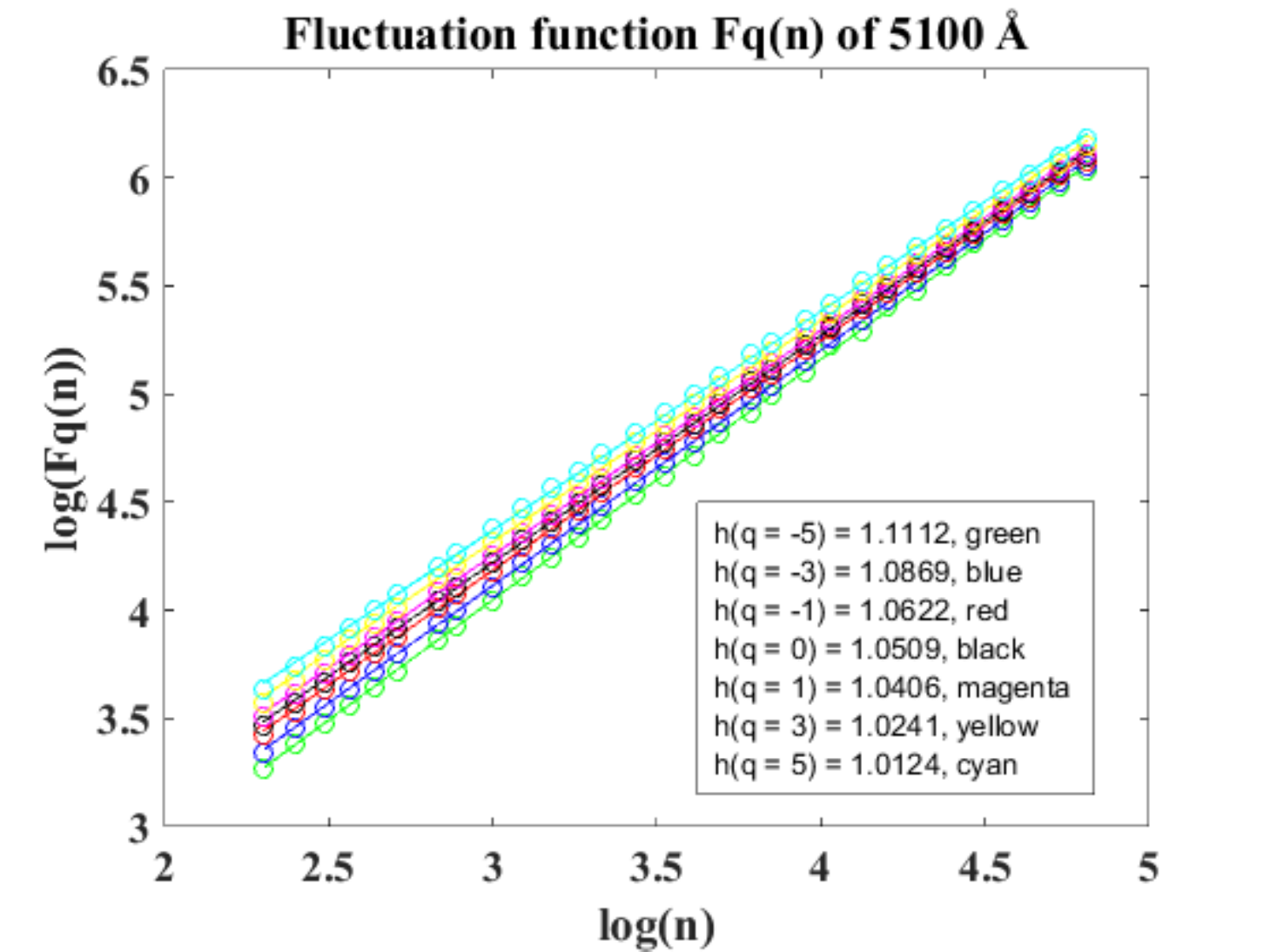}
\caption{Top panel: the fluctuation function $F_q$($n$) of the H$\beta$ emission line in the time segment from day 47508 to day 52175. The one on right side is after least square fitting for selected $q$ values. Bottom panel: the same as the top panel but for the 5100 {\AA} continuum flux.}
\label{fig4}
\end{figure*}

\begin{figure*}
\centering
\includegraphics[scale=0.6]{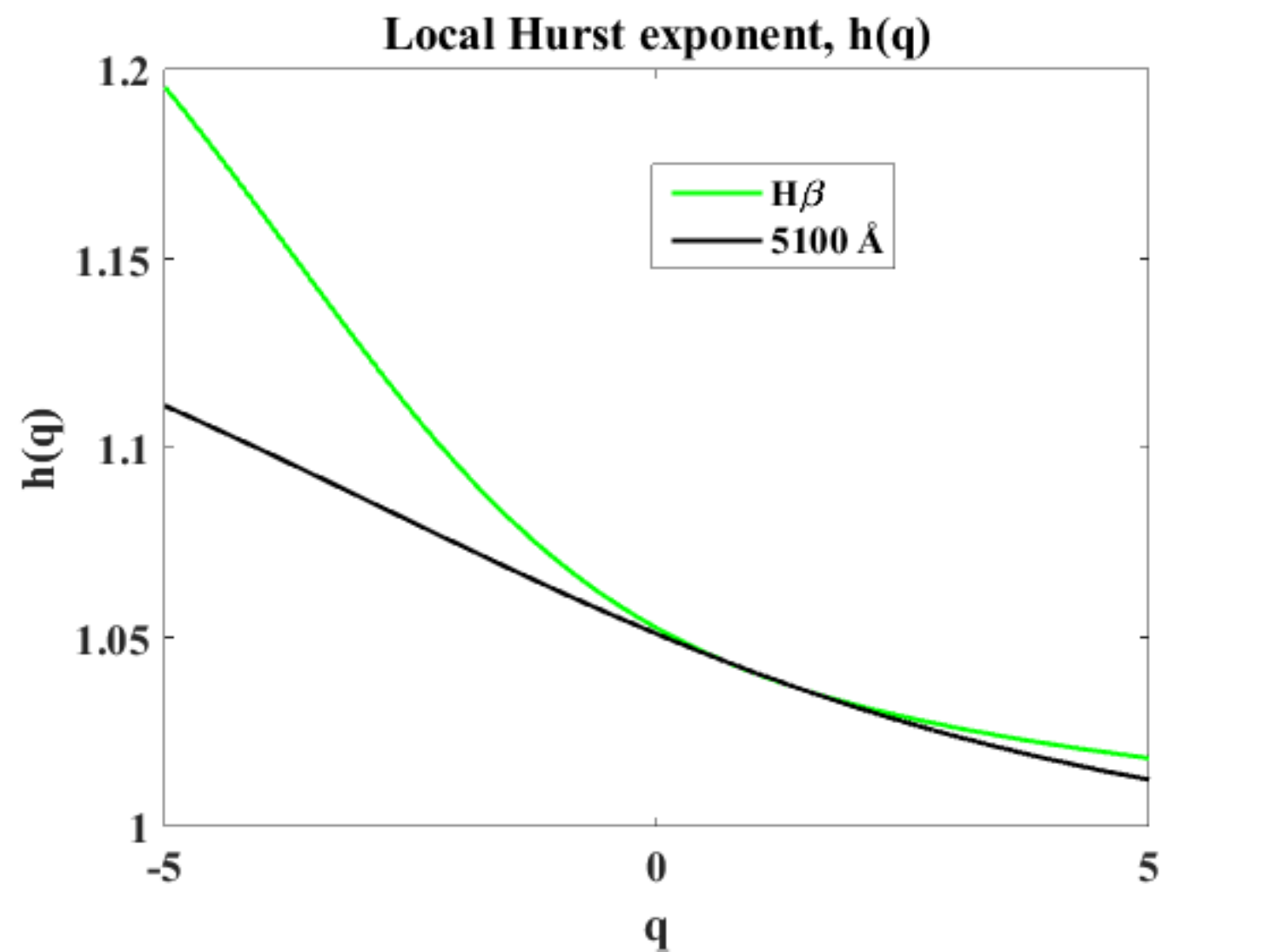}
\includegraphics[scale=0.6]{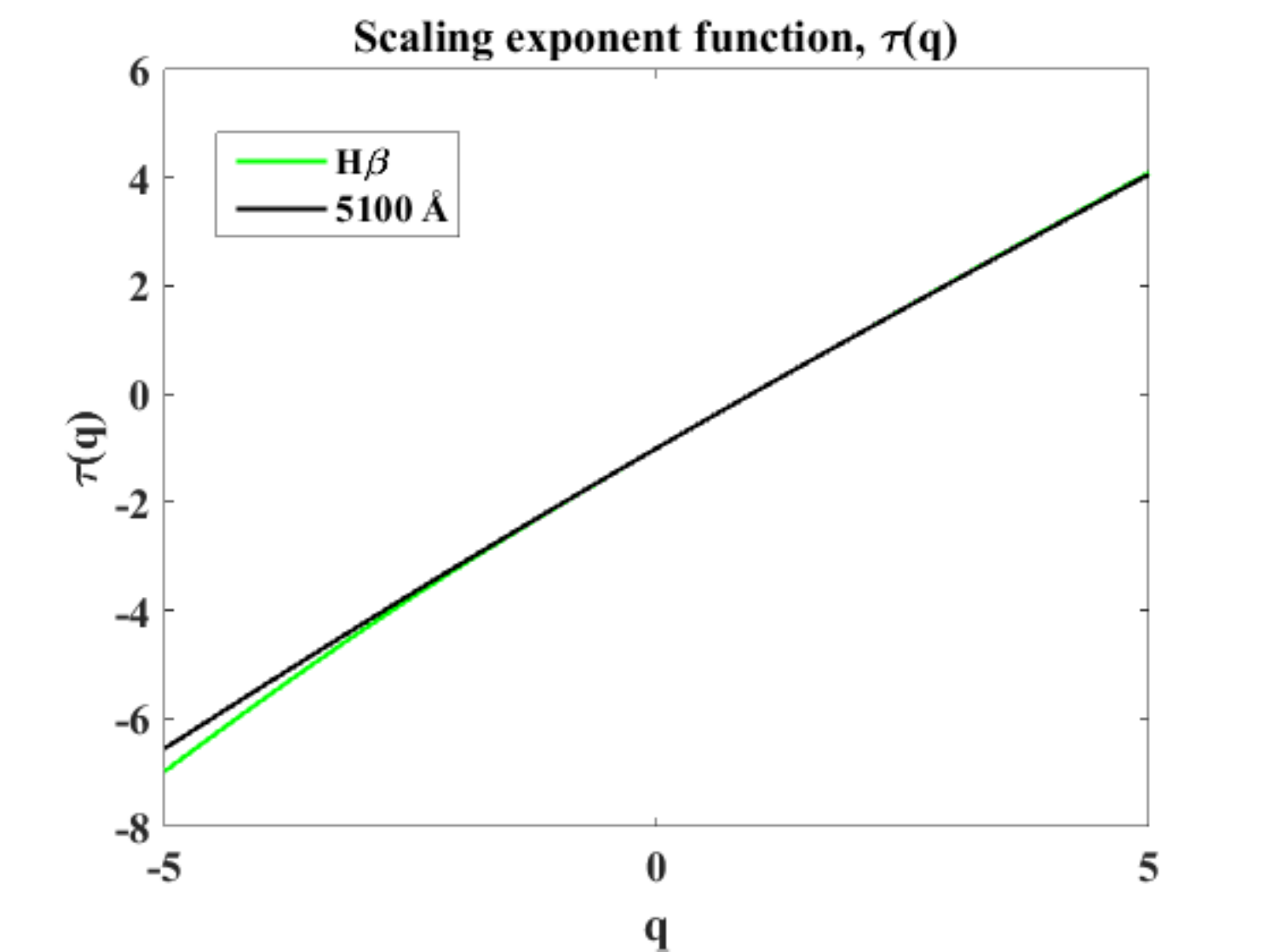}
\includegraphics[scale=0.6]{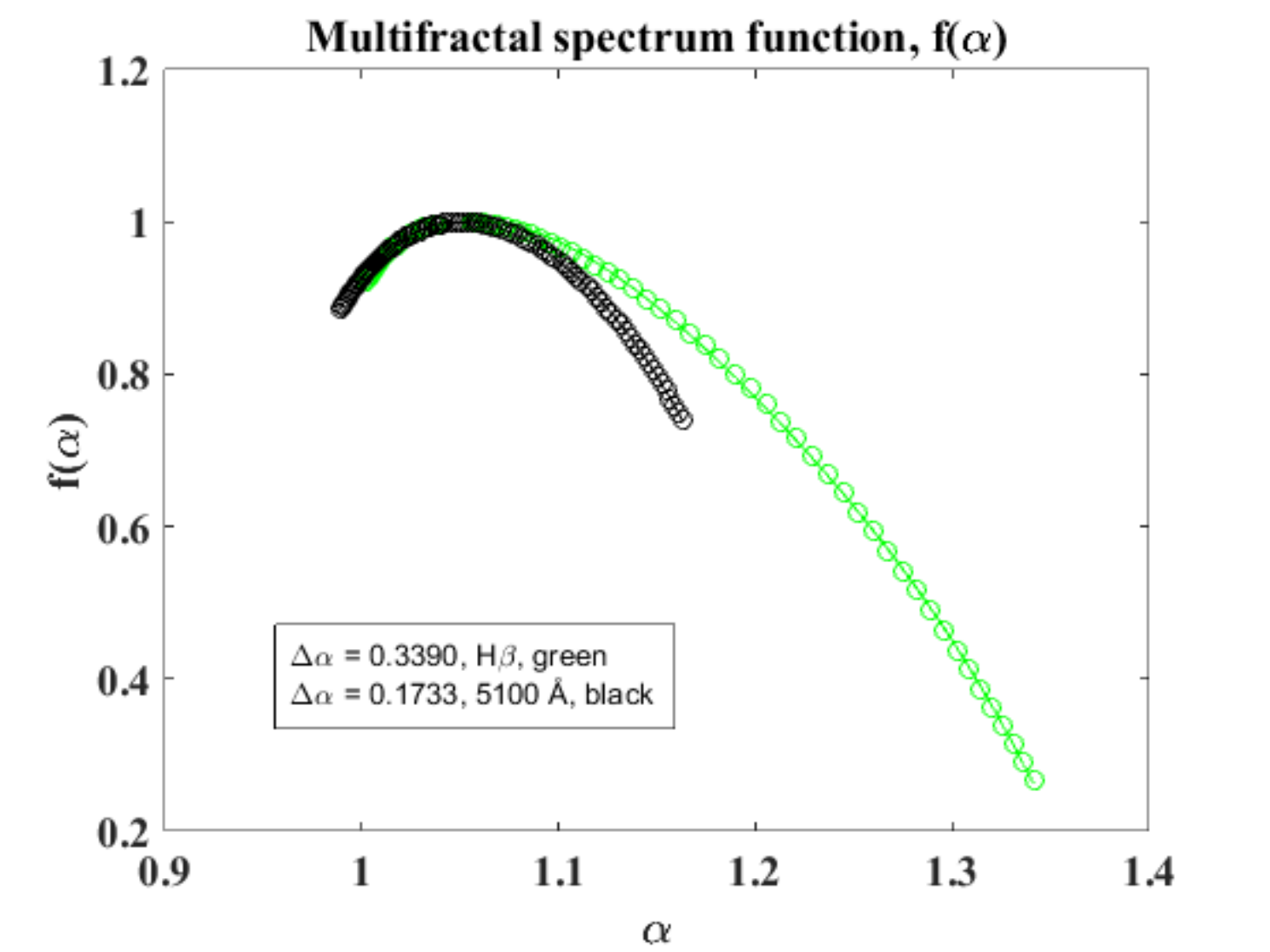}
\caption{Top panel: the local Hurst exponent $h$($q$) (left) and the scaling exponent function $\tau$($q$) (right) for the H$\beta$ emission line (green) and the 5100 {\AA} continuum flux (black) light curves in the time segment from day 47508 to day 52175. Bottom panel: the  multifractal spectrum function $f$($\alpha$) for the H$\beta$ emission line (green) and the 5100 {\AA} continuum flux (black). The circles represent the data points, whereas, the lines represent the fit to fourth-order polynomial.}
\label{fig5}
\end{figure*}

Here we analyze the multifractal signatures present in the light curves of the H$\beta$ line and 5100 {\AA} continuum of NGC 5548 using the MFDMA procedures. First, we determine the scaling relationship between the overall fluctuation function $F_q$($n$) and the scale $n$ for each light curve, Fig. \ref{fig2}, to understand how the fluctuation function scales at each scale $n$. To make sure the nature of relationship existed between the fluctuation function $F_q$($n$) and the scale $n$, we calculate the slope, usually known as the local Hurst exponents $h(q)$, from the log-log plot of $F_q$($n$) versus $n$, for each moment $q$ using the least square fitting technique, Fig. \ref{fig3}. In addition, we determine the classical scaling exponent function $\tau$($q$) using Eq.~\ref{eq10} and present the results obtained for each light curve in Fig. \ref{fig3}. Further more, we estimate the multifractal spectrum function $f$($\alpha$) by using Eqs.~\ref{eq11} and \ref{eq12} for each light curve, Fig. \ref{fig3}. The main purpose of estimating the multifractal spectrum function is mainly to determine the range of exponents present in the light curves and quantify the degree of multifractality detected, which in turn can be used to identify the light curve with strong multifractal signature.\\
Since the data between day 47508 and day 52175 are more densely sampled and of better quality than the rest (\citealt{2016ApJS..225...29B}), we repeat the same analysis for this time segment separately. The results obtained, the fluctuation functions $F_q$($n$), the local Hurst exponents $h$($q$), the Renyi scaling exponents $\tau$($q$), and the multifractality spectrum functions $f$($\alpha$), are given in  Figs. \ref{fig4} and \ref{fig5}, respectively.
\textbf{\section{Discussion}\label{dis}}
The results obtained are discussed as follows. The local Hurst exponents $h$($q$) describes the scaling behaviour of the scales with large and small fluctuations for positive and negative values of $q$, respectively (\citealt{tanna2014multifractality}). For a multifractal time series, the fluctuation function behaves differently for negative and positive values of $q$ at the smallest segment sizes (\citealt{ ihlen2012introduction}). In addition, for a multifractal time series, the  local Hurst exponents $h$($q$) shows nonlinear dependency on the moment $q$ and decreases monotonically when $q$ increases (\citealt{tanna2014multifractality}). In contrast, for monofractal time series, $h$($q$) has a constant value, i.e., the  local Hurst exponents do not change with the moment $q$ (\citealt{kantelhardt2002multifractal}). As shown in Fig. \ref{fig2} and \ref{fig3}, for both light curves, $F_q$($n$) is different for positive and negative values of $q$ at the smallest scales, and the local exponent $h$($q$) has a nonlinear relationship with the moment $q$ and decreases monotonically when $q$ increases, revealing the presence of multifractal signature both in the H$\beta$ line and 5100 {\AA} continuum light curves of NGC 5548. The difference in the nature of the relationship between the fluctuation function $F_q$($n$) and the scale $n$, or the difference in the behaviour of the local Hurst exponents between the light curves indicates the presence of different degree of multifractality. Judging only from the value of the local exponent at the second moment $h$($q$ = 2), this is 1.0255 for $H\beta$ and 1.0308 for 5100 {\AA}, so we are dealing with $h$($q$ = 2) $>$ 1 non-stationary time series. Using the relation $H$ = $h$($q$ = 2) - 1 (\citealt{2006JSMTE..02..003S}), we find that $H$ = 0.0255 ($H\beta$) and $H$ = 0.0308 (5100 {\AA}), indicating the presence of anti-persistent long-range correlations that tends not to continue in the same direction but to turn back on itself giving a less smooth signal (\citealt{Hampson:11}).  In addition, the power spectrum scaling exponent $\delta = 2H+1 = 1.0510$ ($H\beta$) and $\delta = 1.0616$ (5100 {\AA}), which are consistent with random walk like time series or fBm processes (see details in section \ref{sec2}). The measured power-law index of the $PSD$ is a kind of average of the two asymptotic values for the DRW model, and thus, the observed time series may be in rough agreement with DRWs for some choices of $\tau$ and $\delta$. However, we note that this last statement is exclusively based on the second moment $h$($q$ = 2), and one should properly check the DRW predictions for all moments $h$($q$). A multifractal analysis of DRW-based simulated light curves is out of the scope of this paper, although we verified that $SFs$ for both time series over a wide range of time lags have complex behaviours, which cannot be reproduced for DRW models.\\
To verify the discussion based on the fluctuation function and the local Hurst exponents, we determine the scaling exponent function $\tau$($q$). Similarly, for monofractal time series  $\tau$($q$) is independent of $q$ whereas it shows a nonlinear behaviour for a multifractal time series (\citealt{tanna2014multifractality, ihlen2012introduction, kantelhardt2002multifractal}). For the light curves considered here, the nonlinear functionality between the moment $q$ and the scaling exponent function $\tau$($q$) confirms the presence of multifractal signature in the light curves. The difference in the behaviour of $\tau$($q$) between the light curves again shows that the degree of multifractality, and thus nonlinearity between the light curves is different.\\
Once the presence of multiscaling (multifractal) behaviour is confirmed, it is important to determine how strong is the detected multifractal signature. To this end, we calculate the width (singularity strength) of the multifractal spectra by using $\Delta\alpha = \alpha_{max} - \alpha_{min}$. The width value for the H$\beta$ line and 5100 {\AA} continuum are 0.4507 and 0.2002, respectively. Since width of the multifractal spectrum is a measure of degree of multifractality (\citealt{ashkenazy2003nonlinearity}), the wider the width the stronger (richer) the multifractality (\citealt{kantelhardt2002multifractal}). Hence, comparing the width values, the degree of multifractality of the H$\beta$ line is stronger than that of 5100 {\AA}.\\
It has been indicated that the width and shape of the multifractal spectrum are related to temporal variation of the local Hurst exponents. The symmetric spectrum is originated from the leveling of the $q^{th}$  order local Hurst exponent both for negative and positive $q's$ values (\citealt{doi:10.1093/mnras/sty1316} and the references therein). The leveling of $q^{th}$ order local Hurst exponents reflects that the $q$-order fluctuation is insensitive to the magnitude of local fluctuations. A multifractal spectrum will be found with right truncation if the multifractal structure is sensitive to the small-scale fluctuation with large magnitude; whereas, the multifractal spectrum will be found with left truncation if the structure is sensitive to the local fluctuation with small magnitudes. For our case, the multifractal spectra of both the light curves are found to be left-side truncated, indicating that the multifractal structure is sensitive to the local fluctuation with small magnitudes and the existence of small-scale intermittency in the continuum and BLR. In addition, we take into account the effects of the observational noise (flux uncertainties) to the width $\Delta\alpha$ and shape of the spectra. We generated simulated light curves of the continuum and H$\beta$ emissions at epochs equal to those of observation, modifying the observed fluxes by adding random quantities. These additive random numbers were realizations of normal distributions around zero, with standard deviations equal to the measured uncertainties. Our analysis method was then applied to the simulated curves to produce distributions of spectral shapes and widths. Despite the two spectral shapes (H$\beta$ line and 5100 {\AA} continuum) do not change when flux uncertainties are taken into account, we found appreciable differences in the spectral widths. These differences allowed us to infer 1$\sigma$ errors: 0.45 $\pm$ 0.04 (H$\beta$) and 0.20 $\pm$ 0.02 (5100 {\AA}).\\
Similarly, for the data in the time segment between day 47508 and day 52175, the behaviour of all functions, the fluctuation functions, the local Hurst exponents, the Renyi scaling exponents and the multifractal spectra functions, reveal the presence of multifractal, and thus nonlinear signatures in this part of the light curves. The calculated width values for this time segment of the H$\beta$ emission line and the 5100 {\AA} continuum flux light curves are 0.3390 and 0.1733, respectively. Here also the degree of multifractality of the H$\beta$ emission line is stronger than that of 5100 {\AA} continuum. Similarly, to take the flux uncertainty in this particular period into account, we generated simulated light curves of the continuum and H$\beta$ emissions at epochs equal to observation in this time interval. Though the two spectral shapes remain the same when flux uncertainties are taken into account, we found differences in the spectral width. These differences allowed us to infer 1$\sigma$ errors: 0.34 $\pm$ 0.09 (H$\beta$) and 0.17 $\pm$ 0.05 (5100 {\AA}). This may imply that observational noise could affect the width of the multifractal spectrum, but it may not affect the shape of the spectrum. \citet{doi:10.1093/mnras/sty1316} have shown that the widening in the width of a multifractal spectrum could be due to noise contamination.\\
Now, we discuss possible origins of multifractality detected in the light curves. There are two commonly known origins of multifractality: (i) could be due to a broad probability density function for the values of the time series, or/and (ii) due to different long-range correlations for small and large fluctuations (\citealt{kantelhardt2002multifractal}). The simplest way to identify the origin of multifractality in the time series is by analyzing the corresponding shuffled fluxes, i.e., in order to destroy the temporal correlations in the original data, observed fluxes are randomly shuffled at the observing epochs. If the observed multifractality is due to different long-range correlations for small and large fluctuations, then there will be no difference in the fluctuation function $F_q$($n$) of the shuffled data for positive and negative $q's$ at the smallest segment size, and consequently, the local Hurst exponent $h(q)$ will be independent of the moment $q$, having a constant value $h_{shuff}(q)$ = 0.5 at all $q's$. In addition, the scaling exponent $\tau(q)$ of the shuffled data will be a linear function of $q$, and as a result $\Delta\alpha_{shuff}$ = 0. 
\begin{figure*}
\centering
\includegraphics[scale=0.6]{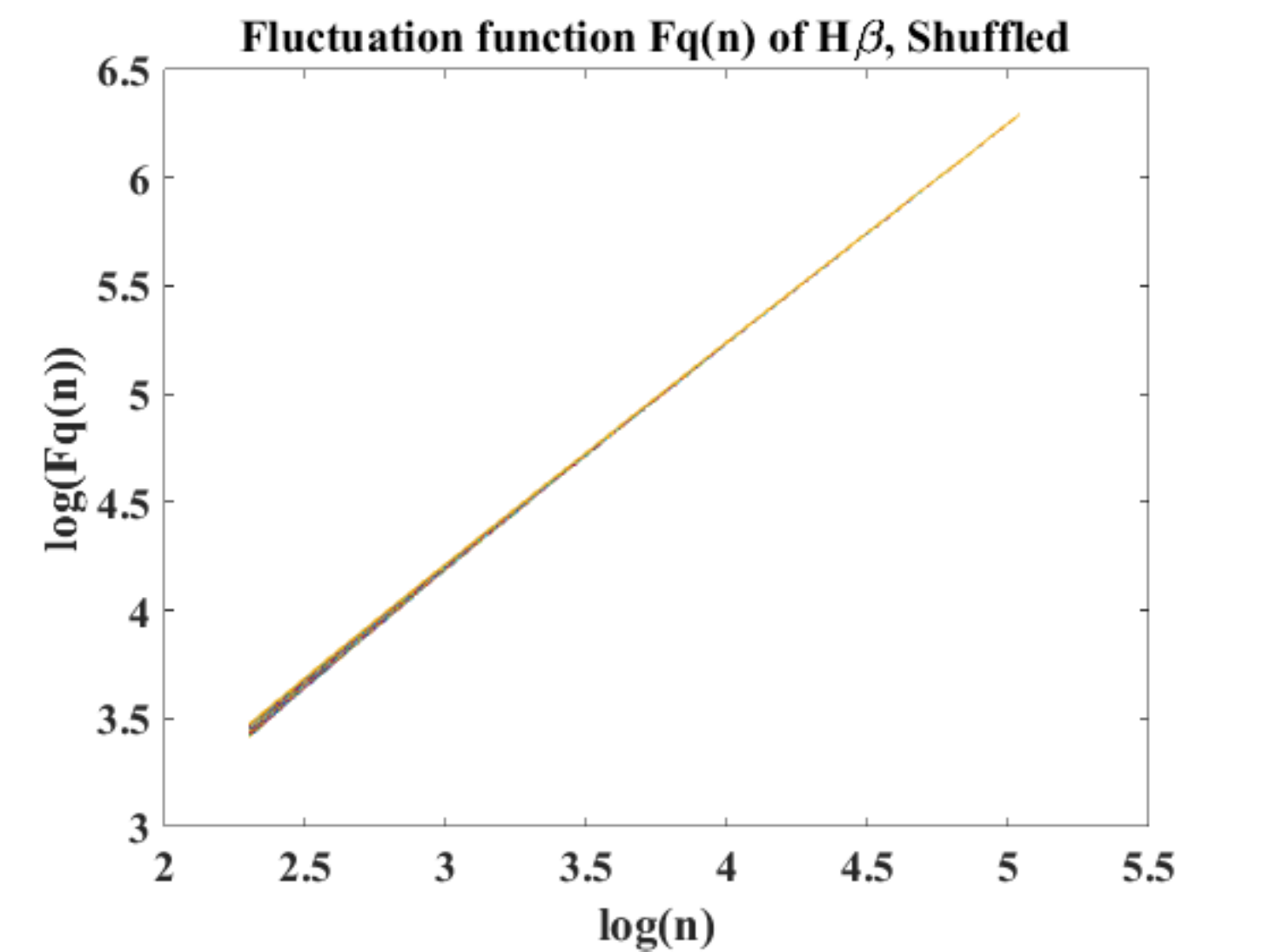}
\includegraphics[scale=0.6]{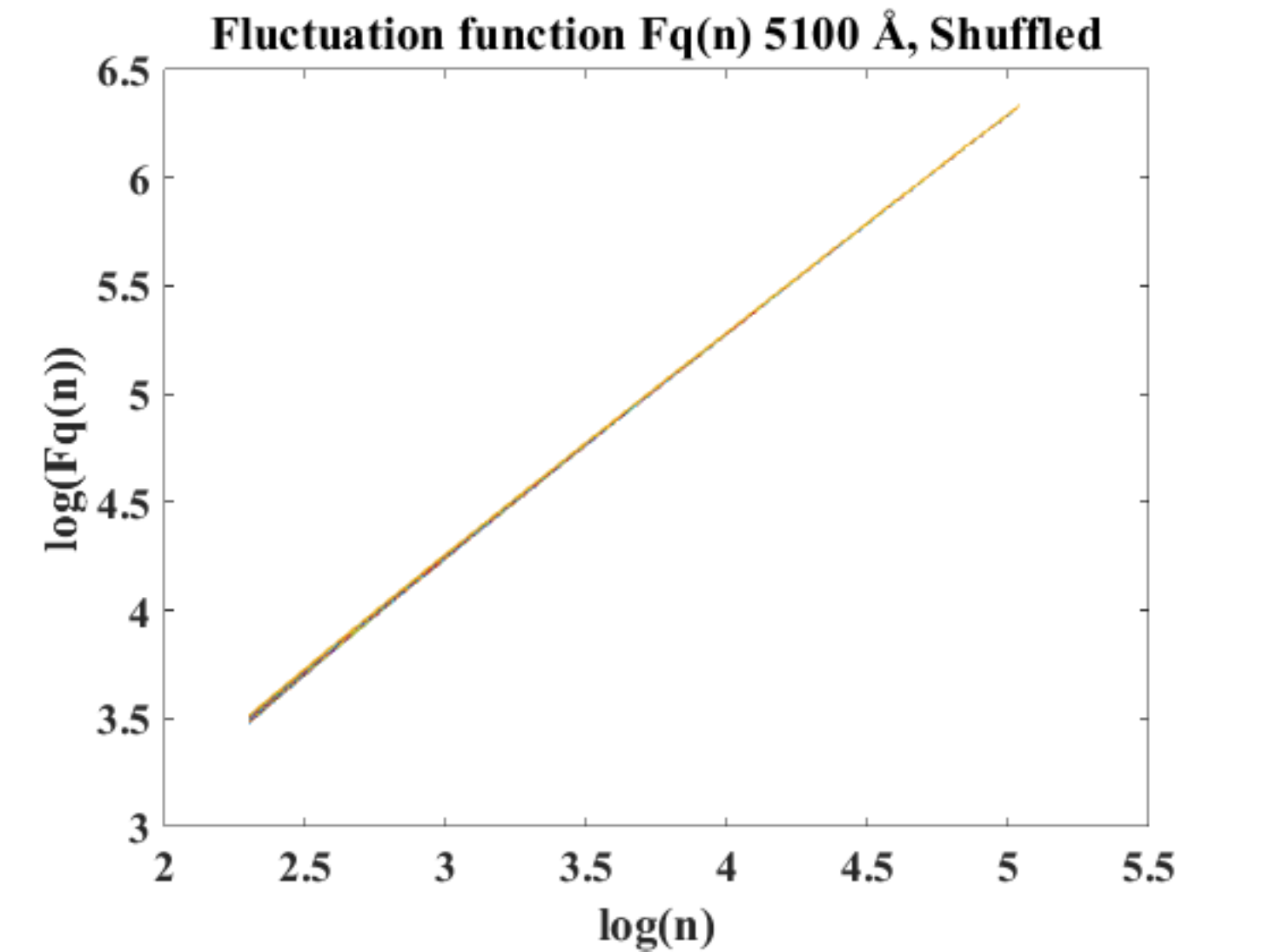}
\caption{The fluctuation function $F_{q}$($n$) of the corresponding shuffled time series of the H$\beta$ emission line (left) and 5100 {\AA} continnum flux light curves.}
\label{fig8}
\end{figure*}
If the multifractality is due to the probability density function, $h_{shuff}$($q$) = $h_{ori}$($q$), and as a result \textbf{$\Delta\alpha_{shuff}$ =  $\Delta\alpha_{ori}$}, this is from the fact that the shuffling process does not affect the probability density function. If both sources of multifractality are present, \textbf{$\Delta\alpha_{shuff}$ $<$ $\Delta\alpha_{ori}$}.
\begin{figure*}
\centering
\includegraphics[scale=0.58]{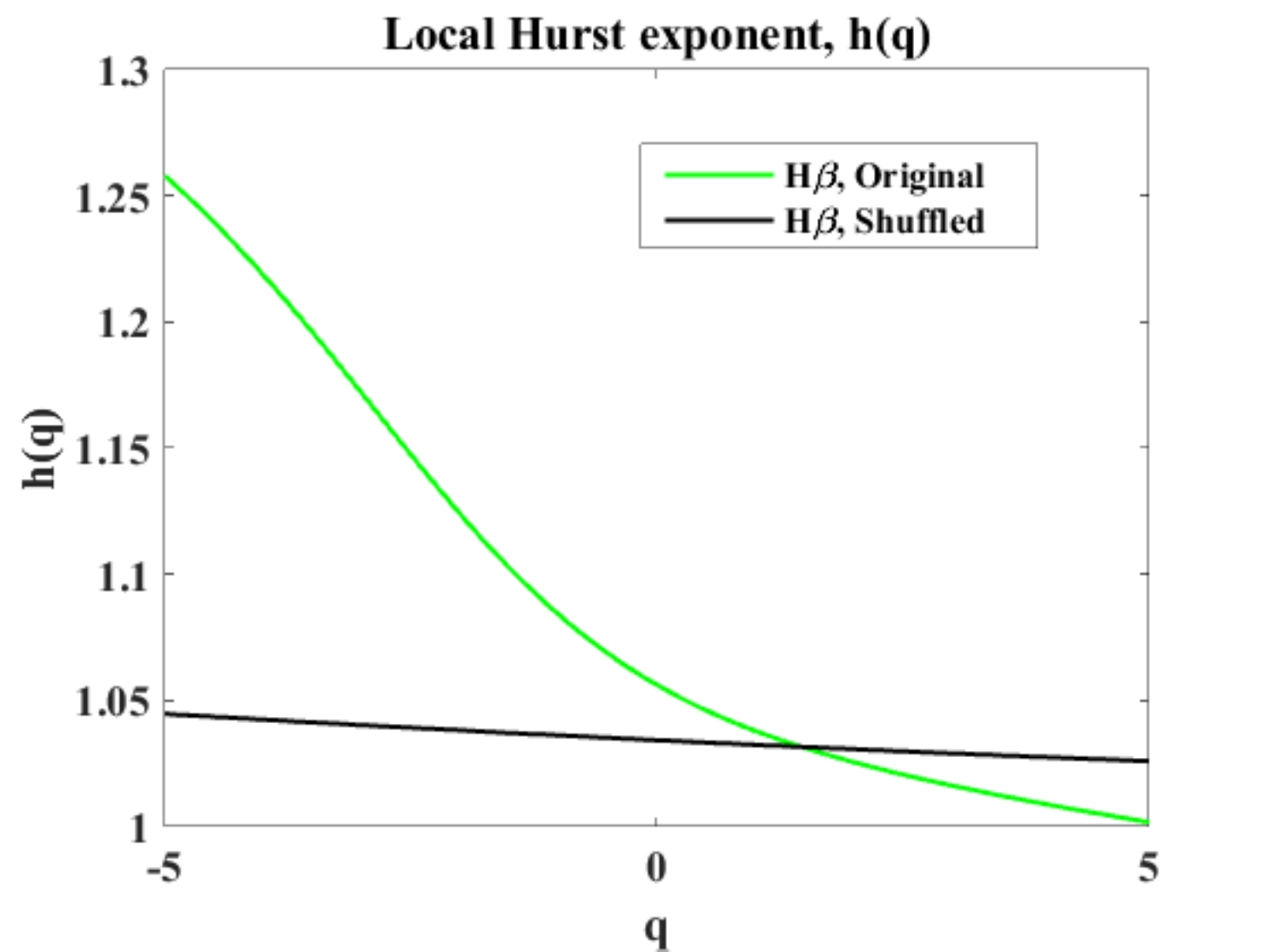}
\includegraphics[scale=0.58]{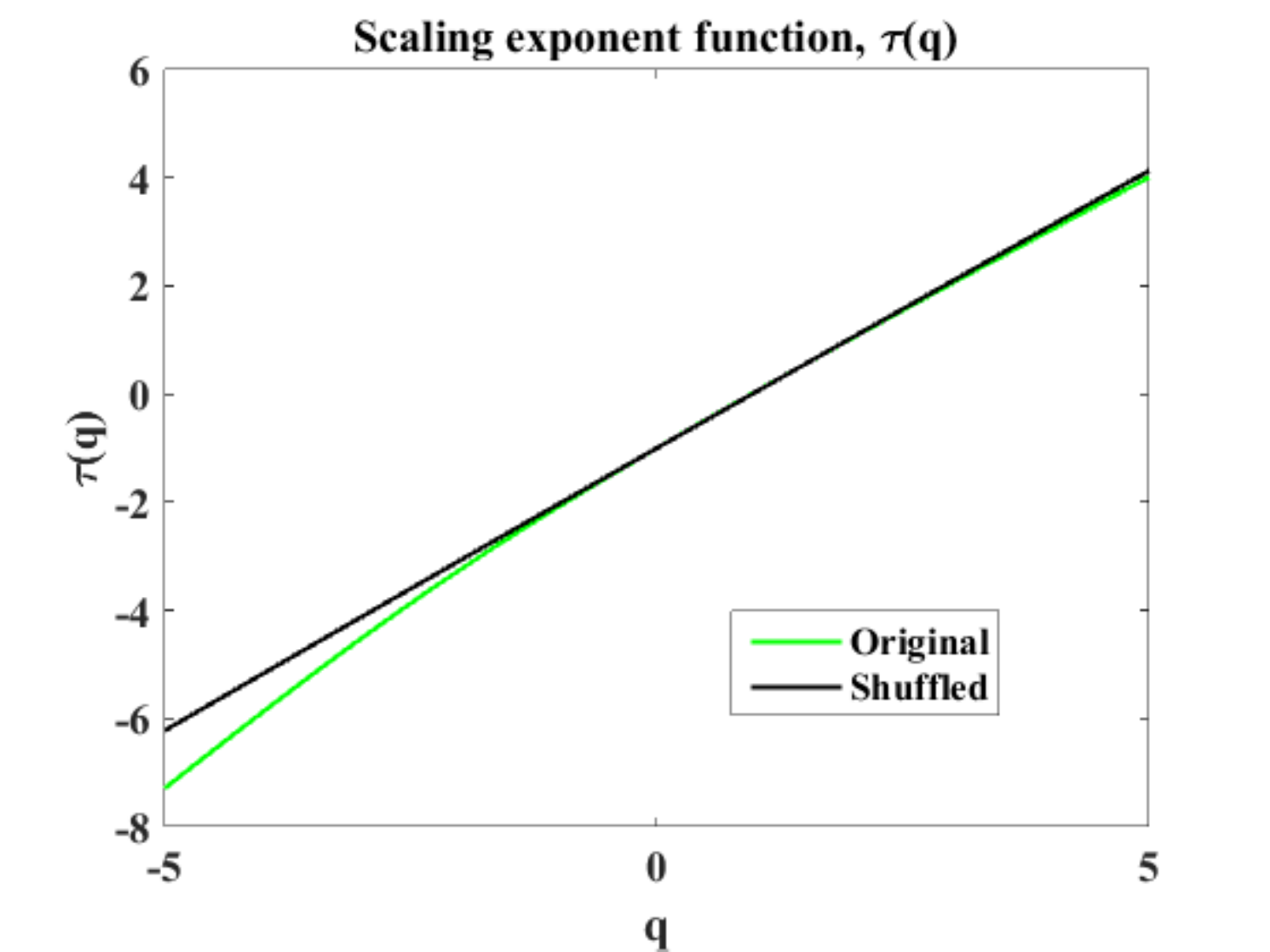}
\includegraphics[scale=0.58]{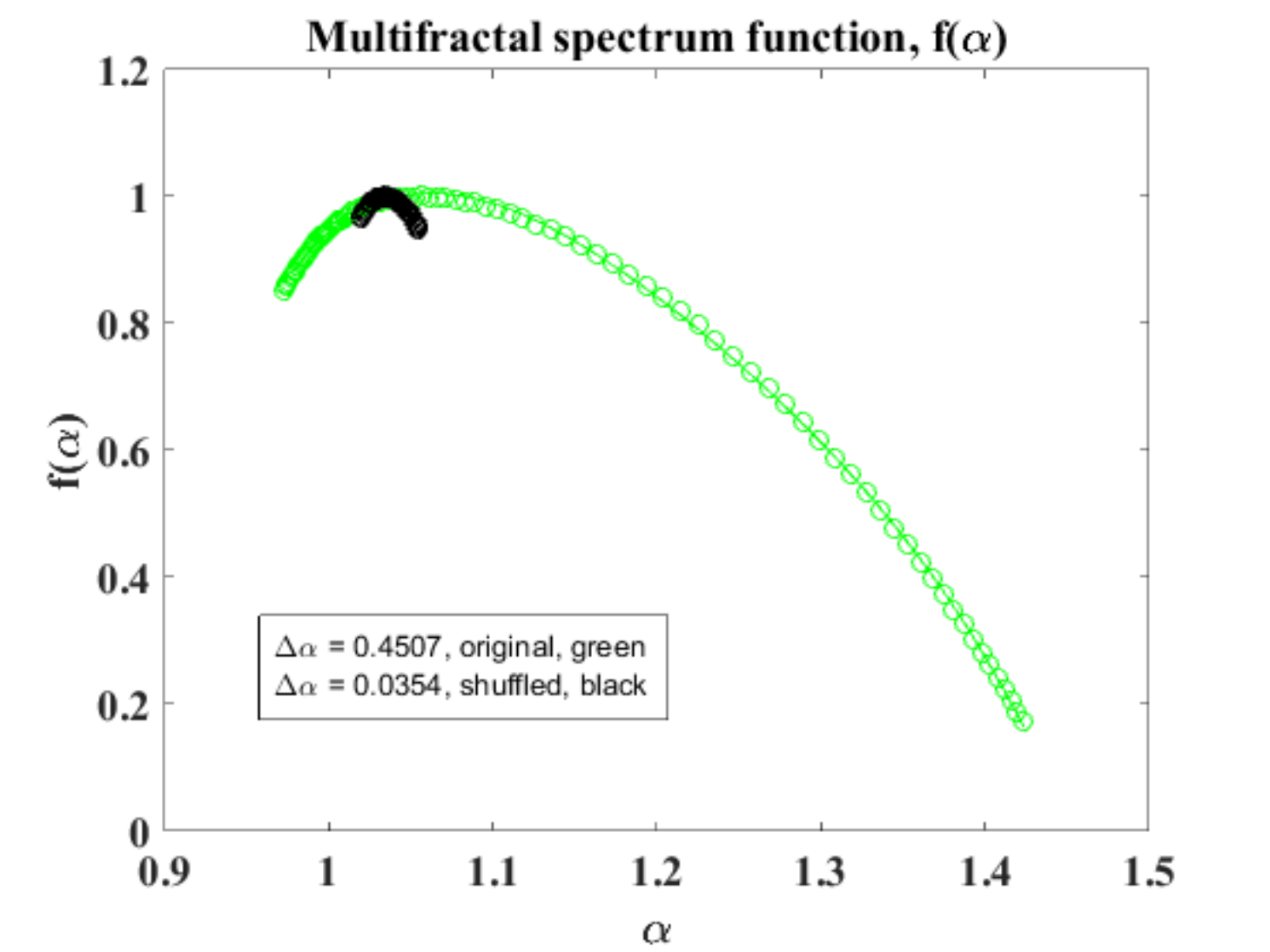}
\caption{Top panel: the local Hurst exponent $h(q)$ (left) and the scaling exponent function $\tau(q)$ (right) for the original light curve of H$\beta$ emission line (green) and the corresponding shuffled time series (black). Bottom panel: the multifractal spectrum function $f(\alpha)$ for the original light curve of the H$\beta$ emission line (green) and the corresponding shuffled time series (black). The circles represent the data points, whereas, the lines represent the fit to fourth-order polynomial.}
\label{fig9}
\end{figure*}
The fluctuation functions $F_{q}(n)$ of the corresponding shuffled time series are given in Fig. \ref{fig8}. In addition, we present the local Hurst exponent $h(q)$, the scaling exponent function $\tau(q)$, and the multifractal spectrum function $f(\alpha)$ of the shuffled series in comparison to the corresponding original light curve of H$\beta$ emission line (Fig. \ref{fig9}) and the 5100 {\AA} continuum (Fig. \ref{fig10}). For the light curves considered here, $h_{shuff}(q)$ is almost constant, but it differs from 0.5. Additionally, $\Delta\alpha_{shuff}$ $<<$ $\Delta\alpha_{ori}$. 
\begin{figure*}
\centering
\includegraphics[scale=0.6]{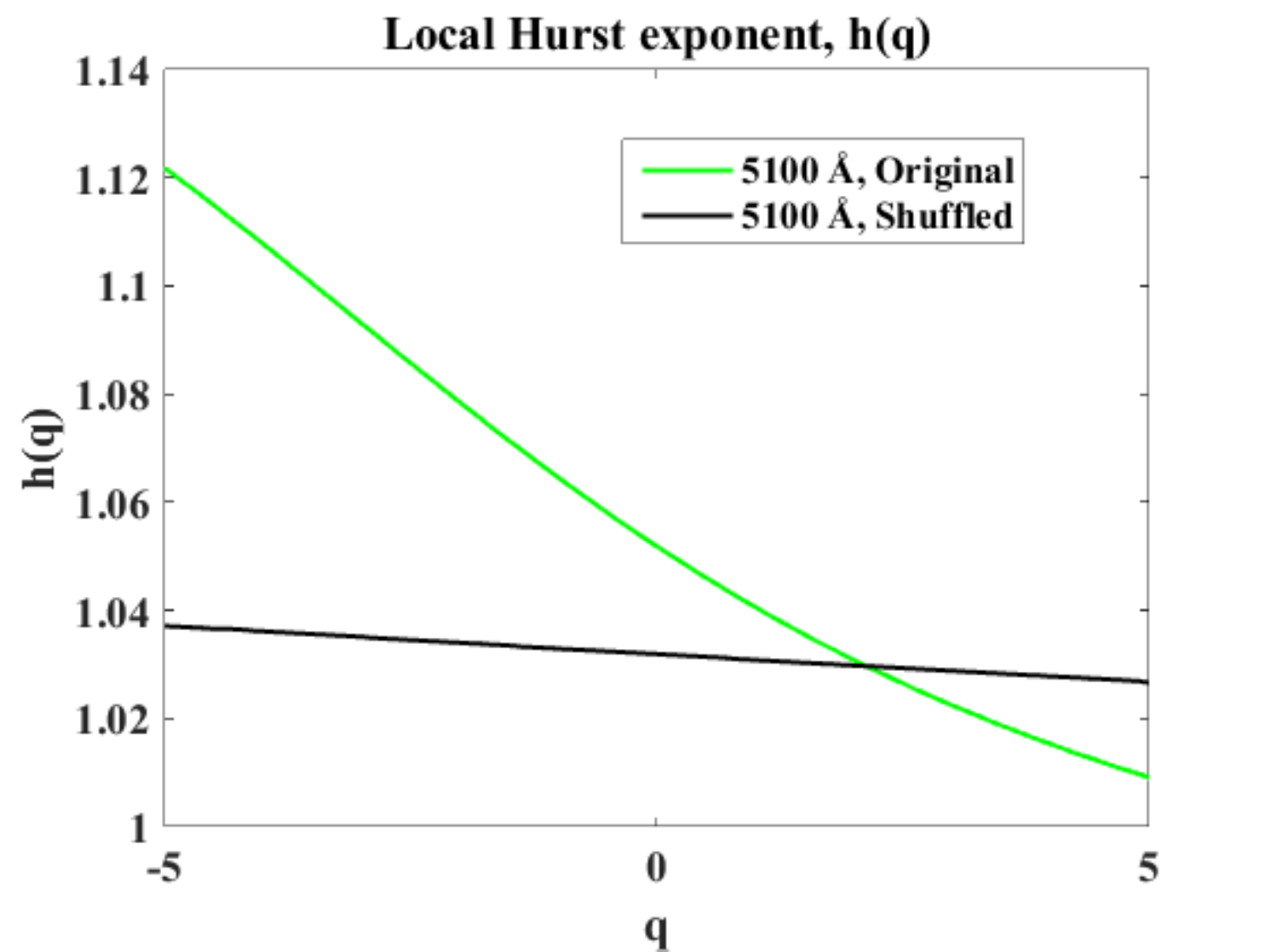}
\includegraphics[scale=0.6]{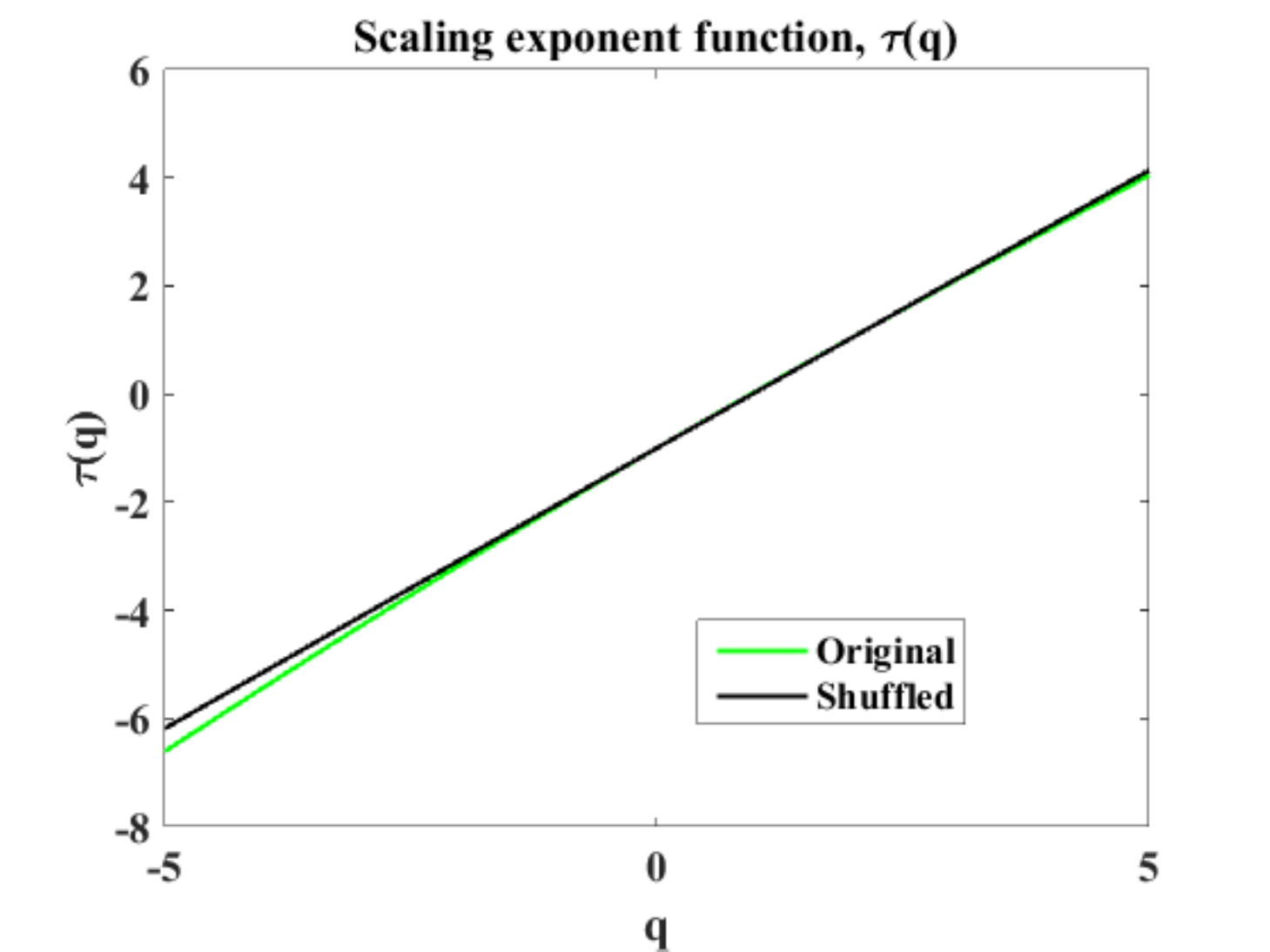}
\includegraphics[scale=0.6]{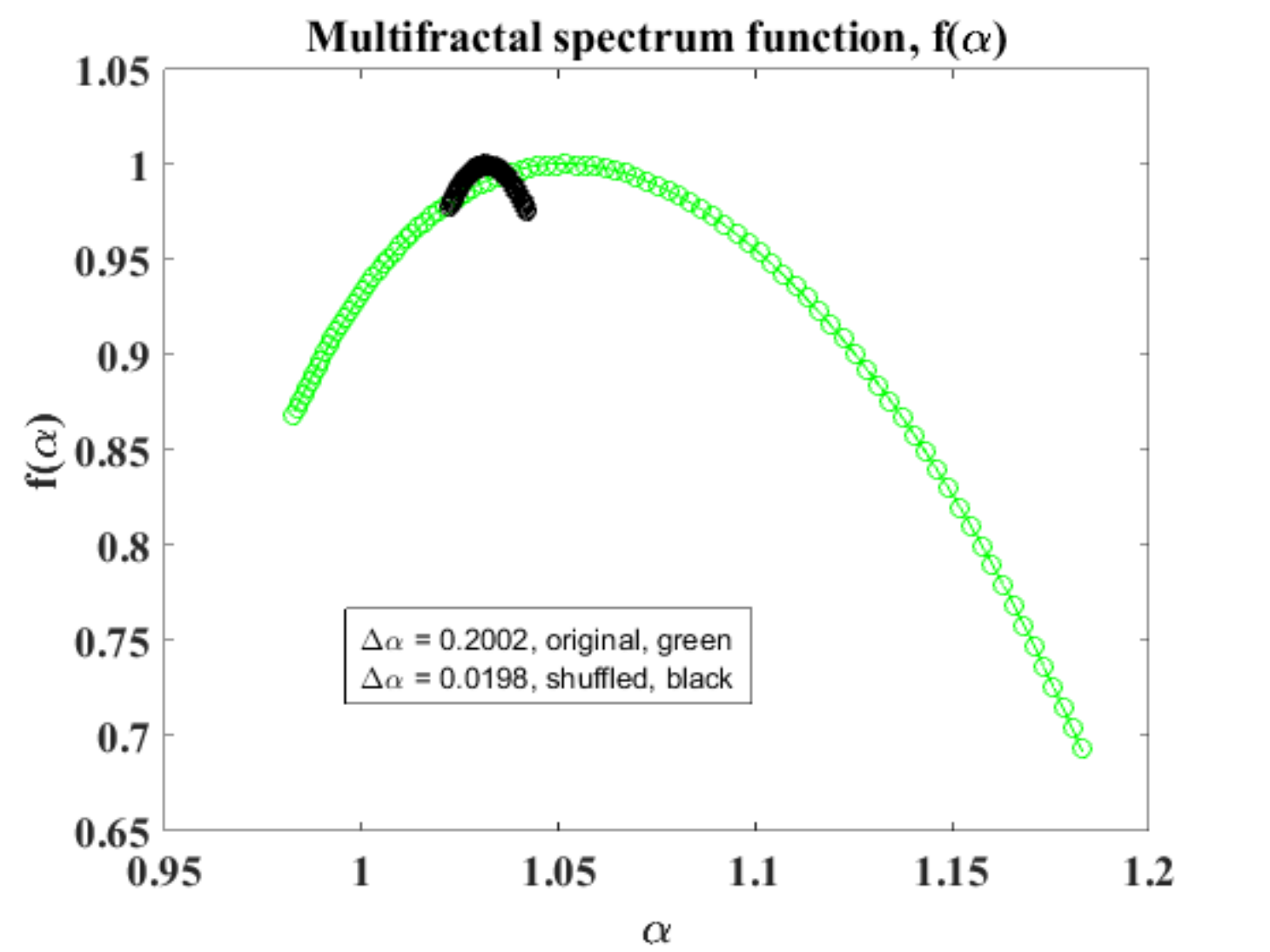}
\caption{Top panel: the local Hurst exponent $h$($q$) (left) and the scaling exponent function $\tau$($q$) (right) for the original light curve of 5100 {\AA} continuum flux (green) and the corresponding shuffled time series (black). Bottom panel: the multifractal spectrum function $f$($\alpha$) for the original light curve of the 5100 {\AA} continuum flux (green) and the corresponding shuffled time series (black). The circles represent the data points, whereas, the lines represent the fit to fourth-order polynomial.}
\label{fig10}
\end{figure*}
Thus, although both mechanisms appear to play a role in the observed multifractality, we find that the long-range correlations contributes more to the observed multifractality, since the shuffling procedure greatly reduced the multifractal signature present in the original time series.  
\section{Summary and Conclusions}\label{concl}
We perform multifractality analysis for the full record and densely sampled period of the H$\beta$ emission line and 5100 {\AA} continuum light curves of the Seyfert type I galaxy NGC 5548  using the backward ($\theta=0$) MFDMA procedures. We also repeat the same analysis for the corresponding shuffled time series to identify the origin of multifractality. First, we calculate the overall fluctuation function $F_q$($n$) for each light curve, from which we estimate the local Hurst exponents $h$($q$) using least square fitting technique. Second, knowing the local Hurst exponents, we determine the Renyi scaling exponent function $\tau$($q$). At last, we estimate the H{\"o}lder exponent $\alpha$($q$) and  multifractal spectra function $f$($\alpha$), from which we calculate the width of the spectra of the light curves. \\
We detect multifractality, and thus nonlinear signatures in the full record and densely sampled period of the H$\beta$ line and 5100 {\AA} continuum light curves of NGC 5548. The H$\beta$ light curves (full record and densely sampled period) exhibit stronger multifractality than those of the 5100 {\AA} continuum i.e., $\Delta\alpha$ (H$\beta$) $\sim$ 2$\Delta\alpha$ (5100 {\AA}). The presence of different anti-persistent long-range temporal correlations for small and large fluctuations is identified to be the main source of the observed multifractal (nonlinear) signatures in both light curves.\\
The detection of multifractal signature in the light curves may indicate the presence of complex and nonlinear interactions in the continuum and H$\beta$ emission regions, which cannot be described by linear models. For example, a period where the BLR weakly responded to changes in the continuum was observed (\citealt{2016ApJ...824...11G}). The observed multifractal signature could be due to turbulence in the continuum and emission line regions, since turbulent dynamics produces multifractal and intermittent structures (e.g., see \citealt{2013PhRvL.110t5002L, 2004AnGeo..22.2431Y}). In general, it is the variation in the flux that results in multifractal structure in a
time series, and therefore the observed multifractality signature could be due to different physical mechanisms. Regarding the optical continuum, its variability may be generated in several ways: AD intrinsic instabilities or flares in a central irradiating source (CIS)
that are reprocessed in the AD (e.g., \citealt{2001ApJ...555..775C}), reprocessing of the CIS-AD (nuclear) variations in the BLR (e.g., \citealt{2001ApJ...553..695K}), and others. \citet{2001ApJ...553..695K} reported that the diffuse continuum emission from the BLR is important for NGC 5548, since the diffuse-to-nuclear continuum ratio is of about 25 \% at 5100 {\AA} (see Fig. 2 in that paper). There is also recent observational evidence for this type of emission in NGC 5548 (e.g., \citealt{2016ApJ...821...56F, 2015ApJ...806..129E}). Thus, signatures of nonlinearity might be produced during reprocessing in the BLR clouds. The nonlinearity of the continuum is less significant than that of the H$\beta$ emission line, which may be plausibly interpreted as a dilution effect, i.e., all the H$\beta$ emission is generated in the BLR after reprocessing of the CIS-AD light, while only a fraction of the continuum comes from BLR clouds.\\
The presence of long-range temporal correlations shows the existence of periodic and/or long timescale physical phenomena. For example, the observed long memory could be due to the mechanism responsible for the $\sim$ 5700 day periodicity in the H$\beta$ emission line and continuum flux light curves of NGC 5548, such as orbiting dusty and dust-free clouds, a binary black hole system, tidal disruption events, or the effect of an orbiting star periodically passing through the AD (\citealt{2016ApJS..225...29B}). Recently, it has been remarked that AGN variability is a complex phenomenon that requires advanced statistical approaches (e.g., \citealt{10.1093/mnras/stv1230}). In addition to several techniques developed to better understand the BLR, the continuum region, and the central engine of AGNs, we strongly believe that multifractality analysis provides a potentially significant information mainly about the nature of variations (the scaling behaviour) in time series of these objects, which can be used to constrain their dynamics and study the evolution of nonlinear trends in different regions.

\section*{Acknowledgements}
This publication makes use of a long-term H$\beta$ emission line and 5100 AA (5100 {\AA}) continuum light curves of NGC 5548 in the VizieR On-line Data Catalog. Research activities of the Observational Astronomy Board of the Federal University of Rio Grande do Norte (UFRN) are supported by continuous grants from CNPq and FAPERN Brazilian agencies. We also acknowledge financial support from INCT INEspa\c{c}o/CNPq/MCT. We warmly thank the anonymous referee for valuable suggestions and comments. AB acknowledges a CAPES PhD fellowship. LJG acknowledges support by the MINECO/AEI/FEDER-UE grant AYA2017-89815-P and the University of Cantabria.



\end{document}